\documentclass[aps,nofootinbib,longbibliography,prd,twocolumn]{revtex4-1}
\usepackage[babel]{csquotes}
\usepackage{graphicx}
\usepackage{amsmath,amssymb}
\usepackage[colorlinks,citecolor=blue,linkcolor=blue,urlcolor=blue]{hyperref}
\usepackage{mathrsfs}
\usepackage{enumerate}
\def\be{\begin{equation}}
\def\ee{\end{equation}}

\usepackage[normalem]{ulem}

\graphicspath{{./Figures/},{./}}

\begin{document}

\title{White Holes as Remnants: A Surprising Scenario for the End of a Black Hole}
\author{Eugenio Bianchi }
\email{ebianchi@gravity.psu.edu}
\affiliation{Institute for Gravitation and the Cosmos, The Pennsylvania State University, University Park, PA 16802, USA\\
Department of Physics, The Pennsylvania State University, University Park, PA 16802, USA}
\author{Marios Christodoulou }
\email{christod.marios@gmail.com}
\affiliation{CPT, Aix-Marseille Universit\'e, Universit\'e de Toulon, CNRS, F-13288 Marseille, France\\
Dept. of Physics, Southern University of Science and Technology, Shenzhen 518055, P. R. China}
\author{Fabio D'Ambrosio}
\email{fabio.dambrosio@gmx.ch}
\affiliation{
\mbox{CPT, Aix-Marseille Universit\'e, Universit\'e de Toulon, CNRS, F-13288 Marseille, France. } 
}
\author{Hal M. Haggard}
\email{haggard@bard.edu}
\affiliation{Physics Program, Bard College, 30 Campus Road, Annondale-On-Hudson, NY 12504, USA,\\
Perimeter Institute for Theoretical Physics, 31 Caroline Street North, Waterloo, ON, N2L 2Y5, CAN}
\author{Carlo Rovelli}
\email{rovelli@cpt.univ-mrs.fr}
\affiliation{
\mbox{CPT, Aix-Marseille Universit\'e, Universit\'e de Toulon, CNRS, F-13288 Marseille, France. } 
}
\date{\small\today}

\begin{abstract}
\noindent Quantum tunneling of a black hole into a white hole provides a model for the full life cycle of a black hole. The white hole acts as a long-lived remnant, solving the black-hole information paradox. The remnant solution of the paradox has long been viewed with suspicion, mostly because remnants seemed to be such exotic objects. We point out that (i) established physics includes objects with precisely the required properties for remnants: white holes with small masses but large finite interiors; (ii) non-perturbative quantum-gravity indicates that a black hole tunnels precisely into such a white hole, at the end of its evaporation. We address the objections to the existence of white-hole remnants, discuss their stability, and show how the notions of entropy relevant in this context allow them to evade several no-go arguments.  A black hole's formation, evaporation, tunneling to a white hole, and final slow decay, form a unitary process that does not violate any known physics. 

\end{abstract}
\maketitle

\section{Introduction}

The conventional description of black hole evaporation is based on quantum field theory on curved spacetime, with the back-reaction on the geometry taken into account via a mean-field approximation \cite{Hawking1974}.  The approximation breaks down before evaporation brings the black hole mass down to the Planck mass ($m_{{Pl}}\!=\!\sqrt{\hbar c/G} \sim$ the mass of a $\frac12$-centimeter hair). To figure out what happens next we need quantum gravity.   

A quantum-gravitational process that disrupts a black hole was studied in \cite{Rovelli2014,Haggard2014,DeLorenzo2016,Christodoulou2016,Christodoulou2018}. It is a conventional quantum tunneling, where classical equations (here the Einstein equations) are violated for a brief interval.  This alters the causal structure predicted by classical general relativity  \cite{Frolov:1979tu,Frolov:1981mz,Stephens1994,modesto2004disappearance,Modesto2006a,Mazur:2004, Ashtekar:2005cj,Balasubramanian:2006,Hayward2006, Hossenfelder:2009fc,Hossenfelder:2010a, frolov:BHclosed, GambiniPullin2014a,GambiniPullin2014b, Bardeen2014,Giddings1992b}, by modifying the dynamics of the local apparent horizon.  As a result, the \emph{apparent} horizon fails to evolve into an \emph{event} horizon.  

Crucially, the black hole does not just `disappear': it tunnels into a white hole \cite{Narlikar1974,HAJICEK2001,Ambrus2005,Olmedo:2017lvt} (from the outside, an object very similar to a black hole), which can then leak out the information trapped inside.   The likely end of a black hole is therefore not to suddenly pop out of existence, but to tunnel to a white hole, which can then slowly emit whatever is inside and disappear, possibly only after a long time \cite{Aharonov1987,Giddings1992a,Callan1992,Giddings1993,Preskill1993,Banks1993,Banks1995, Ashtekar:2008jd,Ashtekar:2010qz,Ashtekar:2010hx,Rama2012,Almheiri:2013wka, Chen:2015,Malafarina:2017}. 

The tunneling probability may be small for a macroscopic black hole, but becomes large toward the end of the evaporation. This is because it  increases as the mass decreases.  Specifically, it will be suppressed at most by the standard tunneling factor 
\be
p\sim e^{-{S_E}/{\hbar}}\label{suppression}
\ee
where $S_E$ is the Euclidean action for the process.  This can be estimated on dimensional grounds for a stationary black hole of mass $m$ to be $S_E\sim G m^2/c$, giving 
\be
p\sim e^{-({m}/{m_{{Pl}}})^2},\label{suppression2}
\ee
which becomes of order unity towards the end of the evaporation, when $m \to m_{{Pl}}$. A more detailed derivation is in \cite{Christodoulou2016,Christodoulou2018}. As the black hole shrinks towards the end of its evaporation, the probability to tunnel into a white hole is no longer suppressed. The transition  gives rise to a long-lived white hole with Planck size horizon and very large but finite interior.  Remnants in the form of geometries with a small throat and a long tail were called ``cornucopions" in \cite{Banks1992} by Banks \emph{et.al.} and studied in \cite{Giddings1992c, Banks1993b,Giddings1994,Banks1995}. As far as we are aware, the connection to the conventional white holes of general relativity was never made. 

This scenario offers a resolution of the information-loss paradox. Since there is an {\em apparent} horizon but no {\em event} horizon, a black hole can trap information for a long time, releasing it after the transition to white hole.  If we have a quantum field evolving on a black hole background metric and we call $S$ its (renormalized) entanglement entropy across the horizon, then consistency requires the metric to satisfy non-trivial conditions: 

(a) The remnant has to store information with entropy $S\sim m_o^2/\hbar$ (we adopt units $G\!\!=\!\!c\!\!=\!\!1$, while keeping $\hbar$ explicit), where $m_o$ is the \emph{initial} mass of the hole, before evaporation \cite{Marolf2017}. This is needed to purify Hawking radiation.
 
(b) Because of its small mass, the remnant can release the inside information only  slowly---hence it must be long-lived. Unitarity and energy considerations impose that its lifetime be equal to or larger than $\tau_R\sim m_o^4/\hbar^{3/2}$ \cite{Preskill1993,Bianchi:2014bma}.

(c) The metric has to be stable under perturbations, so as to guarantee that information can be released \cite{Frolov2012,Barrabes:1993,Poisson:1994, DeLorenzo2016}.

In this paper we show that under simple assumptions the effective metric that describes standard black hole evaporation followed by a transition to a Planck-mass white hole satisfies precisely these conditions.  This result shows that this scenario is consistent with known physics and does not violate unitarity. 

One reason this scenario may not have been recognised earlier is because of some prejudices (including against white holes), which we discuss below.  But the scenario presented here turns out to be  consistent with general expectations that are \emph{both}  in the AdS/CFT community (see for instance \cite{Engelhardt2016, Fitzpatrick2016}) and in the quantum gravity community (see for instance the `paradigm' \cite{Ashtekar:2005cj}).

\section{The internal geometry before quantum gravity becomes relevant}

\begin{figure}[b]
\includegraphics[height=4.5cm]{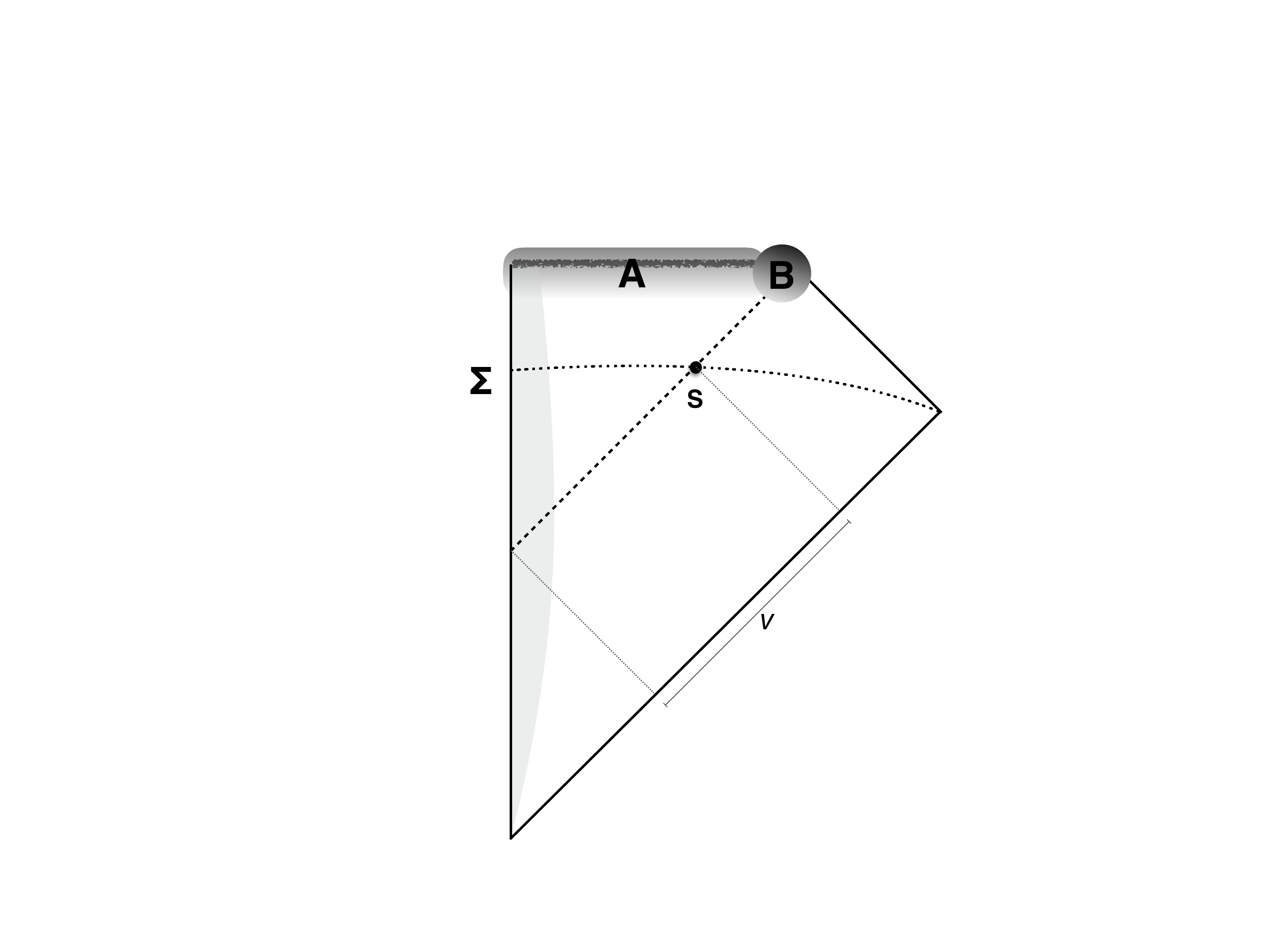}
\caption{\em Conformal diagram of a classical black hole. The dashed line is the horizon. The dotted line is a Cauchy surface $\Sigma$.  In regions $A$ and $B$ we expect (distinct) quantum gravitational effects and classical GR is unreliable.}
\label{uno}
\end{figure}

We begin by studying the geometry \emph{before} any quantum gravitational effect becomes relevant.  The standard classical conformal diagram of a black hole formed by collapsing matter is depicted in Figure \ref{uno}, for the case of spherical symmetry.  

Classical general relativity becomes insufficient when either ($a$) curvature becomes sufficiently large, or ($b$) sufficient time has ellapsed.  The two corresponding regions, $A$ and $B$, where we expect classical general relativity to fail are depicted in the figure.  

Consider the geometry \emph{before} these regions, namely on a Cauchy surface $\Sigma$ that crosses the horizon at some (advanced) time $v$ after the collapse. See Figure \ref{uno}.   We are interested in particular in the geometry of the portion $\Sigma_i$ of $\Sigma$ which is \emph{inside} the horizon.  Lack of focus on this interior geometry is, in our opinion, one of the sources of the current confusion. Notice that we are here fully in the expected domain of validity of  established physics. 

The interior Cauchy surface can be conveniently fixed as follows.  First, observe that a (2d, spacelike) sphere $\mathcal{S}$ in (4d) Minkowski space determines  a preferred (3d) ball $\Sigma_i$ bounded by $\mathcal{S}$: the one sitting on the same linear subspace---simultaneity surface---as $\mathcal{S}$; or, equivalently, the one with maximum volume. (Deformations from linearity in Minkowski space \emph{decrease} the volume).  The first characterisation---linearity---makes no sense on a curved space, but the second---extremized volume---does.  Following \cite{Christodoulou2015}, we use this characterization to fix $\Sigma_i$, which, incidentally, provides an invariant definition of the ``Volume inside $\mathcal{S}$".  Large interior volumes and their possible role in the information paradox have also been considered in \cite{Stanford2014,Perez2015,Ori:2016, AshtekarILQGS:2015,Susskind:2018fmx}.

The interior is essentially a very long tube. As time passes, the radius of the tube shrinks, while its length increases, see Figure 2.

\begin{figure}[h]
\includegraphics[height=3.5cm]{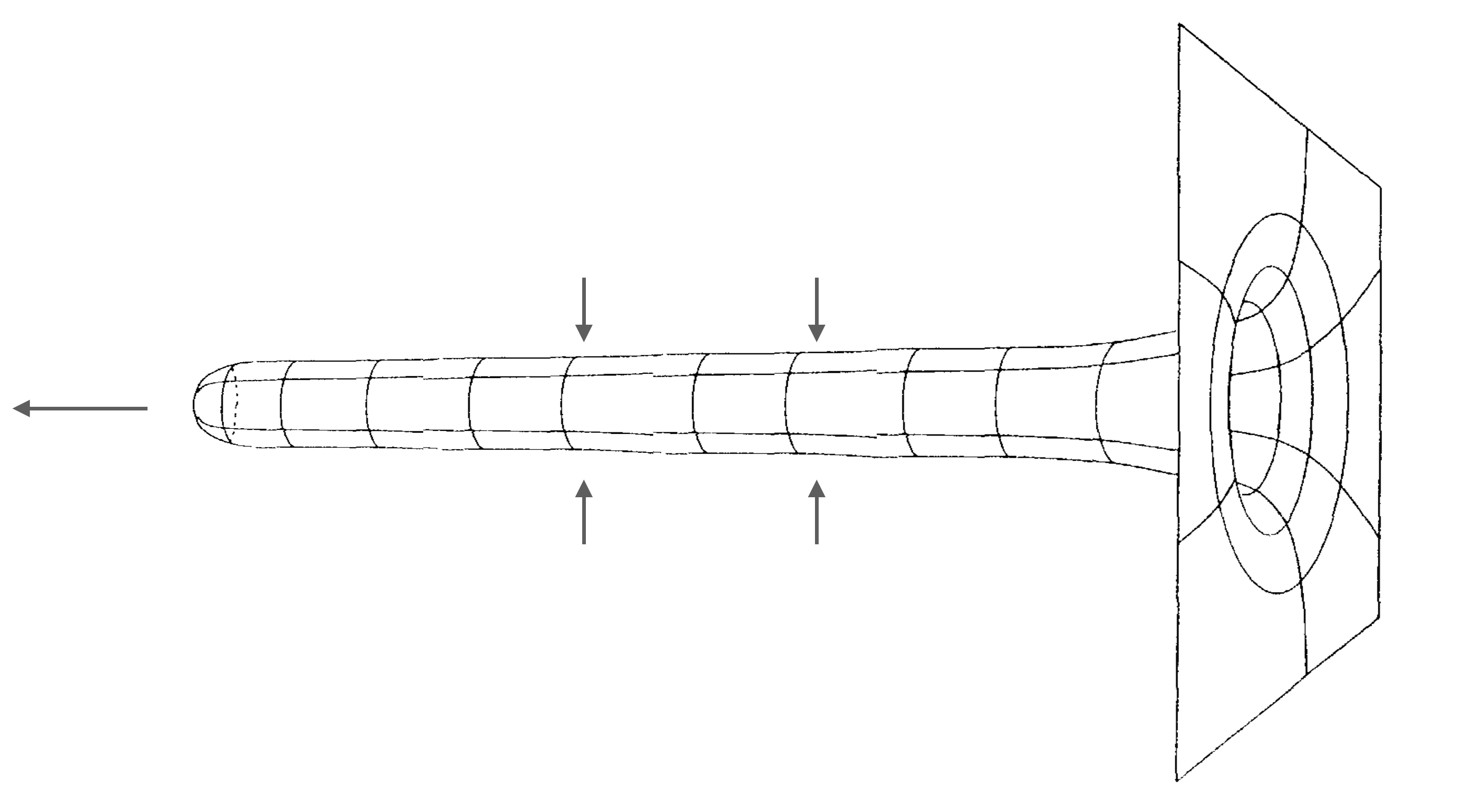}
\caption{\em The interior geometry of an old black hole: a very long thin tube, whose length increases and whose radius decreases with time. Notice it is finite, unlikely the  Einstein-Rosen bridge.}
\label{due}
\end{figure}
It is shown in \cite{Christodoulou2015,Bengtsson2015,Ong2015,Wang2017}, that for large time $v$ the volume of $\Sigma_i$ is proportional to the time from collapse:
\be
        V \sim 3\sqrt{3}\ m_o^2\, v. \label{tre}
\ee 
 Christodoulou and De Lorenzo have shown \cite{Christodoulou2016a} that this picture is not changed by Hawking evaporation: toward the end of the evaporation the area of the (apparent) horizon of the black hole has shrunk substantially, but the length of the interior tube keeps growing linearly with time elapsed from the collapse.  This can be huge for a black hole that started out as macroscopic ($m_o\gg m_{Pl}$), even if the horizon area and mass have become small.  The key point is that \eqref{tre} still hold, with $m_o$ being the \emph{initial} mass of the hole \cite{Christodoulou2016a}, see also \cite{Ong:2015}.

The essential fact that is often neglected, generating confusion, is that an old black hole that has evaporated down to mass $m$ has the same exterior geometry as a young black hole with the same mass, {\em but not the same interior}:  an old, largely evaporated hole has an interior vastly bigger than a young black hole with the same mass.  This is conventional physics. \\

To understand the end of a black hole's evaporation, it is important to distinguish the phenomena concerning the two regions $A$ and $B$ where classical general relativity becomes unreliable. 

Region $A$ is characterised by large curvature and covers the singularity. According to classical general relativity the singularity never reaches the horizon. (N.B.: Two lines meeting at the boundary of a conformal diagram does \emph{not} mean that they meet in the physical spacetime.)  

Region $B$, instead, surrounds the end of the evaporation, which involves the horizon, and affects what happens outside the hole.  Taking evaporation into account, the area of the horizon shrinks progressively until reaching region $B$.  

The quantum gravitational effects in regions $A$ and $B$ are distinct, and confusing them is a source of misunderstanding. Notice that a generic spacetime region in $A$ is \emph{spacelike separated} and in general \emph{very distant} from region $B$.  By locality, there is no reason to expect these two regions to influence one another.   

The quantum gravitational physical process happening at these two regions must be considered separately.   

\section{The $A$ region: Transitioning Across the Singularity} 

To study the $A$ region, let us focus on an arbitrary finite portion of the collapsing interior tube.  As we approach the singularity, the Schwarzschild radius $r_s$, which is a temporal coordinate inside the hole, decreases and the curvature increases.  When the curvature approaches Planckian values, the classical approximation becomes unreliable. Quantum gravity effects are expected to bound the curvature \cite{Narlikar1974,Frolov:1979tu,Frolov:1981mz,Stephens1994,modesto2004disappearance,Mazur:2004, Ashtekar:2005cj, Balasubramanian:2006,Hayward2006, Hossenfelder:2009fc,Hossenfelder:2010a,frolov:BHclosed, Rovelli2013d, Bardeen2014,Giddings1992a,Giddings1992b,Yonika:2017qgo,Olmedo:2017lvt}. Let us see what a bound on the curvature can yield. Following \cite{DAmbrosio}, consider the line element 
\be
ds^2=-\frac{4(\tau^2+l_{})^2}{2m-\tau^2}d\tau^2+\frac{2m-\tau^2}{\tau^2+l_{}}dx^2+(\tau^2+l_{})^2d\Omega^2, \label{me2}
\ee 
where $l_{}\!\ll\!m$. This line element defines a genuine Riemannian spacetime, with no divergences and no singularities. Curvature is bounded. For instance, the Kretschmann invariant $K\equiv R_{\mu\nu\rho\sigma}R^{\mu\nu\rho\sigma}$ is easily computed to be
\begin{eqnarray}
K(\tau)&\approx& \frac{9\,  l_{}^2-24\,  l_{} \tau^2+ 48\, \tau^4}{  (l_{} + \tau^2)^8}m^2
\label{boundedDive}
\end{eqnarray}
in the large mass limit, which has the \emph{finite} maximum  
\be
K(0)\approx\frac{9\, m^2}{ l_{}^6}. \label{K}
\ee

For all the values of $\tau$ where $ l_{}\ll \tau^2 < 2m$ the line element is well approximated by taking $l_{}=0$ which gives 
\be
ds^2=-\frac{4\tau^4}{2m-\tau^2}d\tau^2+\frac{2m-\tau^2}{\tau^2}dx^2+\tau^4d\Omega^2. \label{me}
\ee 
For $\tau<0$, this is the Schwarzschild metric inside the black hole, as can be readily seen going to Schwarzschild coordinates
\be
t_s=x, \ \quad \text{and} \quad \ r_s=\tau^2. 
\ee
For $\tau>0$, this is the Schwarzschild metric inside a \emph{white} hole.   Thus the metric \eqref{me2} represents a continuous transition of the geometry of a black hole into the geometry of a white hole, across a region of Planckian, but bounded curvature. 

Geometrically, $\tau=constant$ (space-like) surfaces foliate the interior of a black hole. Each of these surfaces has the topology $\mathcal{S}^2 \times \mathbb{R}$, namely is a long cylinder.  As time passes, the radial size of the cylinder shrinks while the axis of the cylinder gets stretched.  Around $\tau=0$ the cylinder reaches a minimal size, and then smoothly bounces back and starts increasing its radial size and shrinking its length.  The cylinder  never reaches zero size but bounces at a small finite radius $l_{}$.  The Ricci tensor vanishes up to terms $O(l_{}/m)$.

\begin{figure}[t]
\includegraphics[height=4cm]{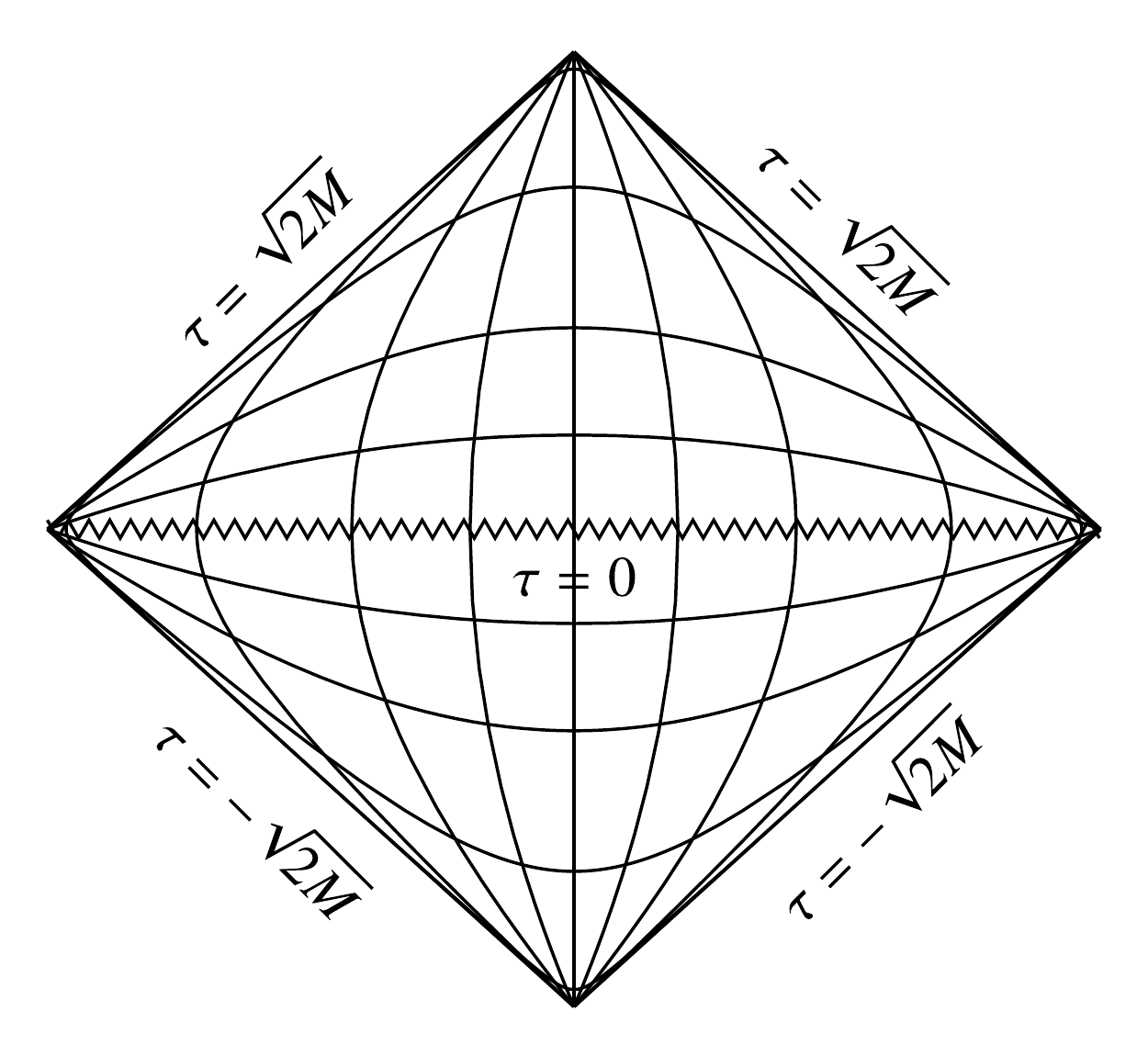}
\caption{\small \em The transition across the $A$ region.}
\label{4}
\end{figure}

The resulting geometry is depicted in Figure \ref{4}. The region around $\tau=0$ is the smoothing of the central black hole singularity at $r_s=0$.

This geometry can be given a simple physical interpretation.   General relativity is not reliable at high curvature, because of quantum gravity.  Therefore the ``prediction" of the singularity by the classical theory has no ground.  High curvature induces quantum particle creation, including gravitons, and these can have an effective energy momentum tensor that back-reacts on the classical geometry, modifying its evolution.  Since the energy momentum tensor of these quantum particles can violate energy conditions (Hawking radiation does), the evolution is not constrained by Penrose's singularity theorem.  Equivalently, we can assume that the expectation value of the gravitational field will satisfy \emph{modified} effective quantum corrections that alter the classical evolution. The expected scale of the correction is the Planck scale.  As long as $l_{}\ll m$ the correction to the classical theory is negligible in all regions of small curvature; as we approach the high-curvature region the curvature is suppressed with respect to the classical evolution, and the geometry continues smoothly past $\tau=0$.

One may be tempted to take $l$ to be Planckian  $l_{Pl}=\sqrt{\hbar G/c^3}\sim\sqrt{\hbar}$, but this would be wrong.  The value of $l$ can be estimated from the requirement that the curvature is bounded at the Planck scale, $K(0) \sim 1/\hbar^2$. Using this in \eqref{K} gives  
\be
\label{lscale}
l\sim (m\,\hbar)^{\frac13},
\ee 
or, restoring for a moment physical units
\be
{l}\sim  l_{Pl}  \left(\frac{m}{m_{Pl}}\right)^{\frac13},
\ee
which is much larger than the Planck length when $m\gg m_{Pl}$ \cite{Rovelli2014}.  The three-geometry inside the hole at the transition time is 
\be
ds_3^2=\frac{2m}{l}dx^2+l^2d\Omega^2. \label{me3}
\ee 
The volume of the ``Planck star" \cite{Rovelli2014}, namely the minimal radius surface is 
\be
V=4\pi l^2\, \sqrt{\frac{2m}{l}}\,(x_{max}-x_{min}). 
\ee
The range of $x$ is determined by the lifetime of the hole from the collapse to the onset of region $B$, as $x=t_s$.  If region $B$ is at the end of the Hawking evaporation, then $(x_{max}-x_{min})\sim m^3/\hbar$ and from Eq. \eqref{lscale}, $l \sim (m \hbar)^{1/3}$, leading to an internal volume at crossover that scales as
\be
V \sim m^4/\sqrt{\hbar}.  \label{V}
\ee
We observe that in the classical limit the interior volume diverges, but quantum effects make it finite. \\

The $l\to 0 $ limit of the line element \eqref{me2} defines a metric space which is a Riemannian manifold almost everywhere and which can be  taken as a solution of the Einstein's equations that is not everywhere a Riemannian manifold \cite{DAmbrosio}.   Geodesics  of this solution crossing the singularity are studied in \cite{DAmbrosio}: they are well behaved at $\tau=0$ and they cross the singularity in a \emph{finite} proper time.  The possibility of this natural continuation of the Einstein equations across the central singularity of the Schwarzschild metric has been noticed repeatedly by many authors. To the best of our knowledge it was first noticed by Synge in the fifties \cite{Synge1950} and rediscovered by Peeters, Schweigert and van Holten in the nineties~\cite{Peeters:1994jz}. A similar observation has recently been made in the context of cosmology in~\cite{Koslowski:2016hds}.  

As we shall see in the next section, what the $\hbar \to 0$ limit does is to confine the transition inside an event horizon, making it invisible from the exterior.  Reciprocally, the effect of turning $\hbar$ on is to de-confine the interior of the hole. 

\section{The transition and the global structure} 

The physics of the $B$ region concerns gravitational quantum phenomena that can happen around the horizon after a sufficiently long time. The Hawking radiation provides the upper bound $\sim m_o^3/\hbar$ for this time. After this time the classical theory does not work anymore.  Before studying the details of the $B$ region, let us consider what we have so far.  

\begin{figure}[h]
\includegraphics[height=3.8cm]{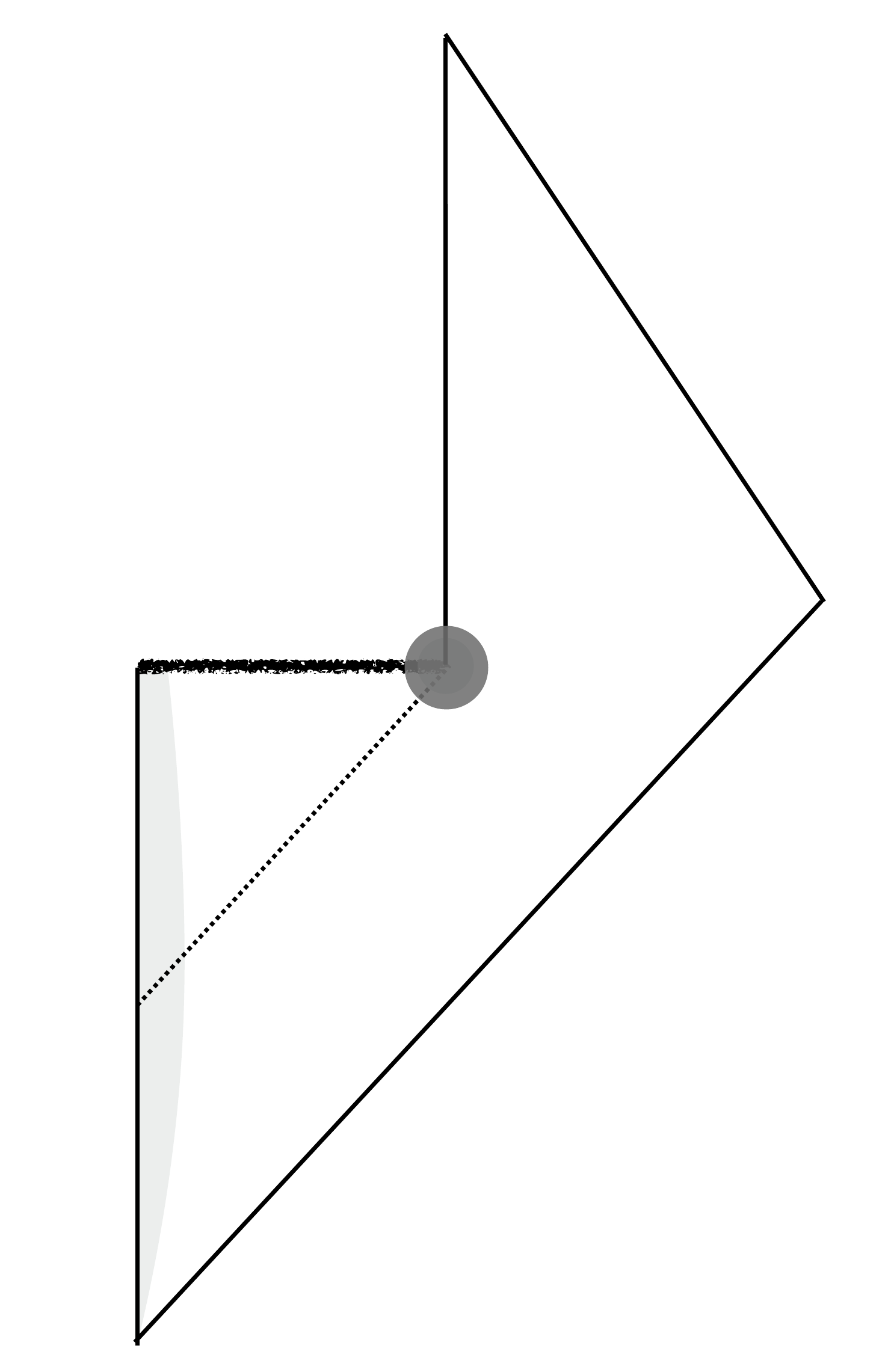}\hspace{.3cm}  \raisebox{1.6cm}{$\Rightarrow$}\hspace{.7cm} \raisebox{-2mm}{\includegraphics[height=4.1cm]{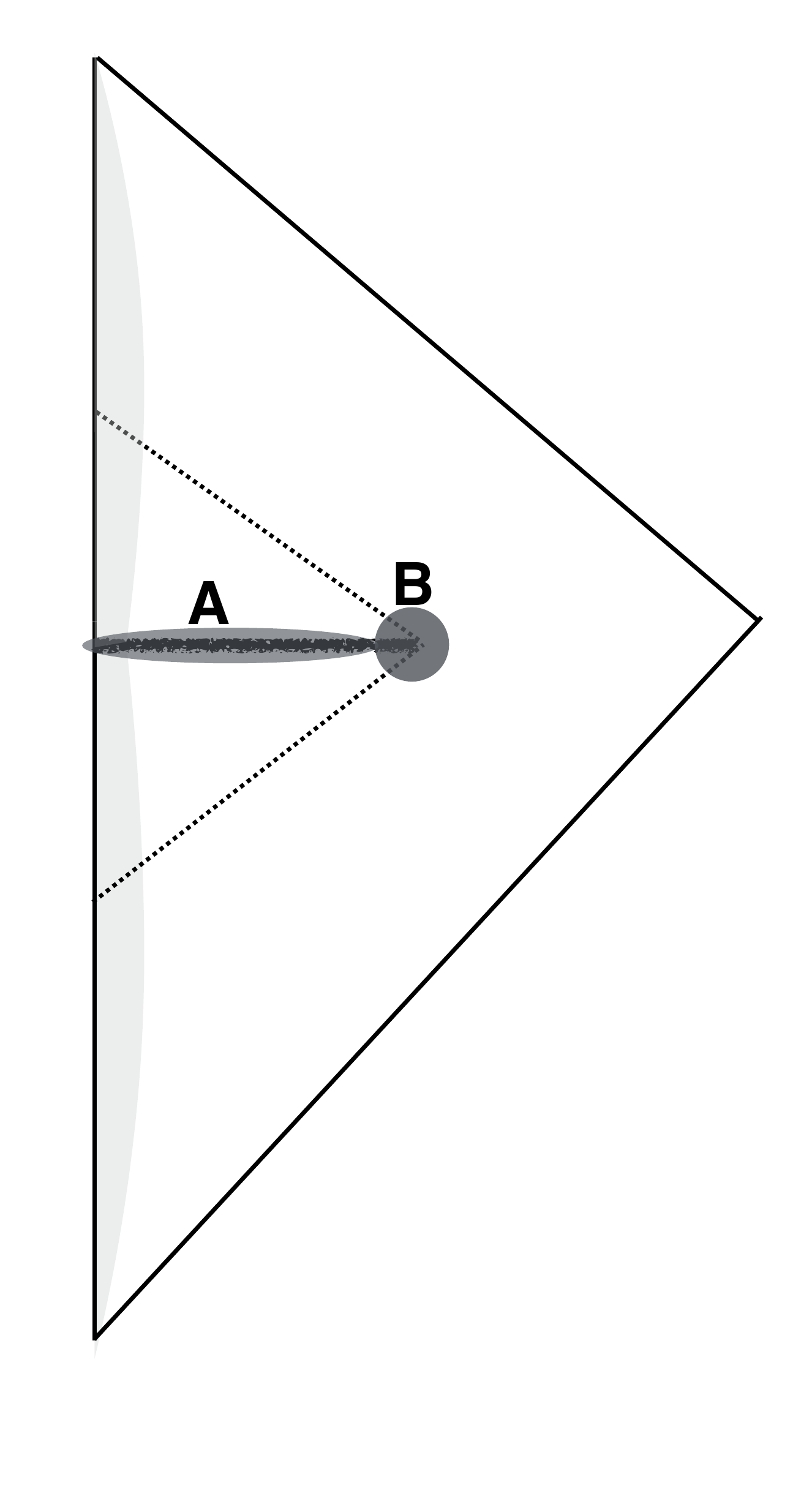}}
\caption{\em Left: A commonly drawn diagram for black hole evaporation that we argue against. Right: A black-to-white hole transition. The dashed lines are the horizons.}
\label{7}
\end{figure}

The spacetime diagram utilized to discuss the black hole evaporation is often drawn as in the left panel of Figure \ref{7}. What happens in the circular shaded region? What physics determines it? This diagram rests on an unphysical assumption:  that the Hawking process proceeds beyond the Planck curvature at the horizon and pinches off the large interior of the black hole from the rest of spacetime. This assumption uses quantum field theory on curved spacetimes beyond its regime of validity.   Without a physical mechanism for the pinching off, this scenario is unrealistic.

Spacetime diagrams representing the possible formation and full evaporation of  a black hole more realistically abound in the literature \cite{Narlikar1974,Frolov:1979tu,Frolov:1981mz,Stephens1994,modesto2004disappearance,Mazur:2004, Ashtekar:2005cj, Balasubramanian:2006,Hayward2006, Hossenfelder:2009fc,Hossenfelder:2010a,frolov:BHclosed, Bardeen2014,Giddings1992a,Giddings1992b} and they are all similar.  In particular, it is shown in \cite{Haggard2014,DeLorenzo2016} that the spacetime represented in the right panel of Figure \ref{7}, can be an {\em exact solution of the Einstein equations}, except for the two regions $A$ and $B$, but including regions within the horizons. 

If the quantum effects in the region $A$ are simply the crossing described in the previous section, this determines the geometry of the region past it, and shows that the entire problem of the end of a black hole reduces to the quantum transition in the region $B$.  

The important point is that there are \emph{two} regions inside horizons: one below and one above the central singularity.  That is, the black hole does not simply pop out of existence: it tunnels into a region that is screened inside an (anti-trapping) horizon. Since it is anti-trapped, this region is actually the interior of a \emph{white} hole. Thus, black holes die by tunneling into white holes.  

Unlike for the case of the left panel of Figure  \ref{7}, now running the time evolution backwards makes sense: the central singularity is screened by an horizon (`time reversed cosmic censorship') and the overall backward evolution behaves qualitatively (not necessarily quantitively, as initial conditions may differ) like the time-forward  one. 

Since we have the explicit metric across the central singularity, we know the features of the resulting white hole.  The main consequence is that its interior is what results from the transition described in the above section: namely a white hole born possibly with a small horizon area, but in any case with {\em a very large interior volume}, inherited from the black hole that generated it.  

\begin{figure}[h]
\includegraphics[height=7cm]{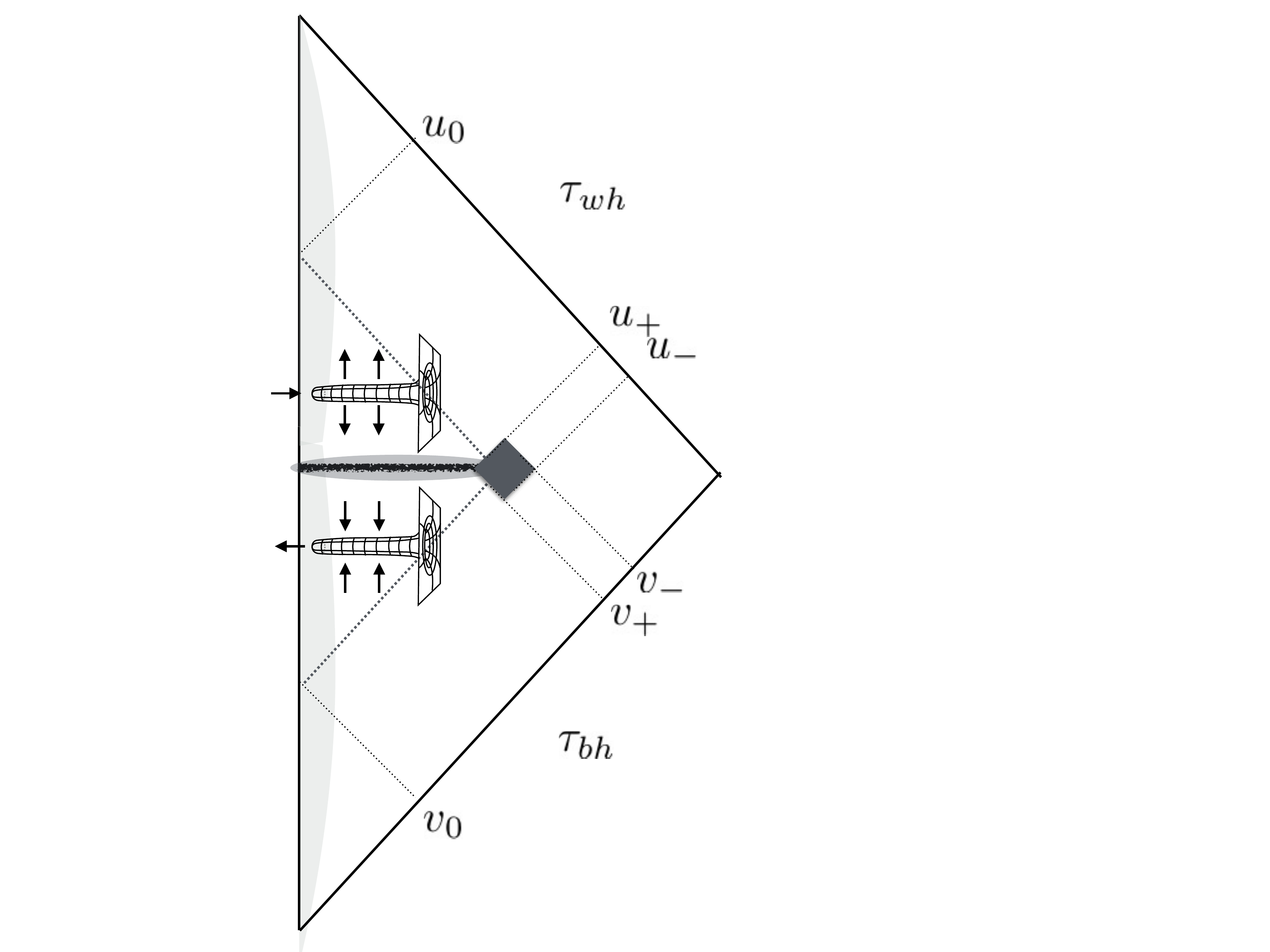}
\vspace{-1em}
\caption{\em Black hole bounce, with a sketch of the inside geometries, before and after the quantum-gravitational transition.}
\label{9}
\end{figure}

If the original black hole is an old hole that started out with a large mass $m_o$, then its interior is a very long tube. Continuity of the size of the tube in the transition across the singularity, results in a white hole formed by the bounce, which initially also consists of a very long interior tube, as in Figure \ref{9}. Subsequent evolution shortens it (because the time evolution of a white hole is the time reversal of that of a black hole), but this process can take a long time. Remarkably, this process results in a white hole that has a small Planckian mass and a long life determined by how old the parent black hole was. In other words, the outcome of the end of a black hole evaporation is a long-lived remnant. 

The time scales of the process can be labelled as in Figure \ref{9}. We call $v_o$ the advanced time of the collapse, $v_-$ and $v_+$ the advanced time of the onset and end of the quantum transition, $u_o$ the retarded time of the final disappearance of the white hole, and $u_-$ and $u_+$ the retarded times of the onset and end of the quantum transition. The black hole lifetime is
\be
    \tau_{bh}=v_--v_o.
\ee
The white hole lifetime is
\be
    \tau_{wh}=u_o-u_+.
\ee
And we assume that the duration of the quantum transition of the $B$ region satisfies $u_+-u_- = v_+-v_-\equiv \Delta \tau$.  

\begin{figure}[h]
\includegraphics[height=3cm]{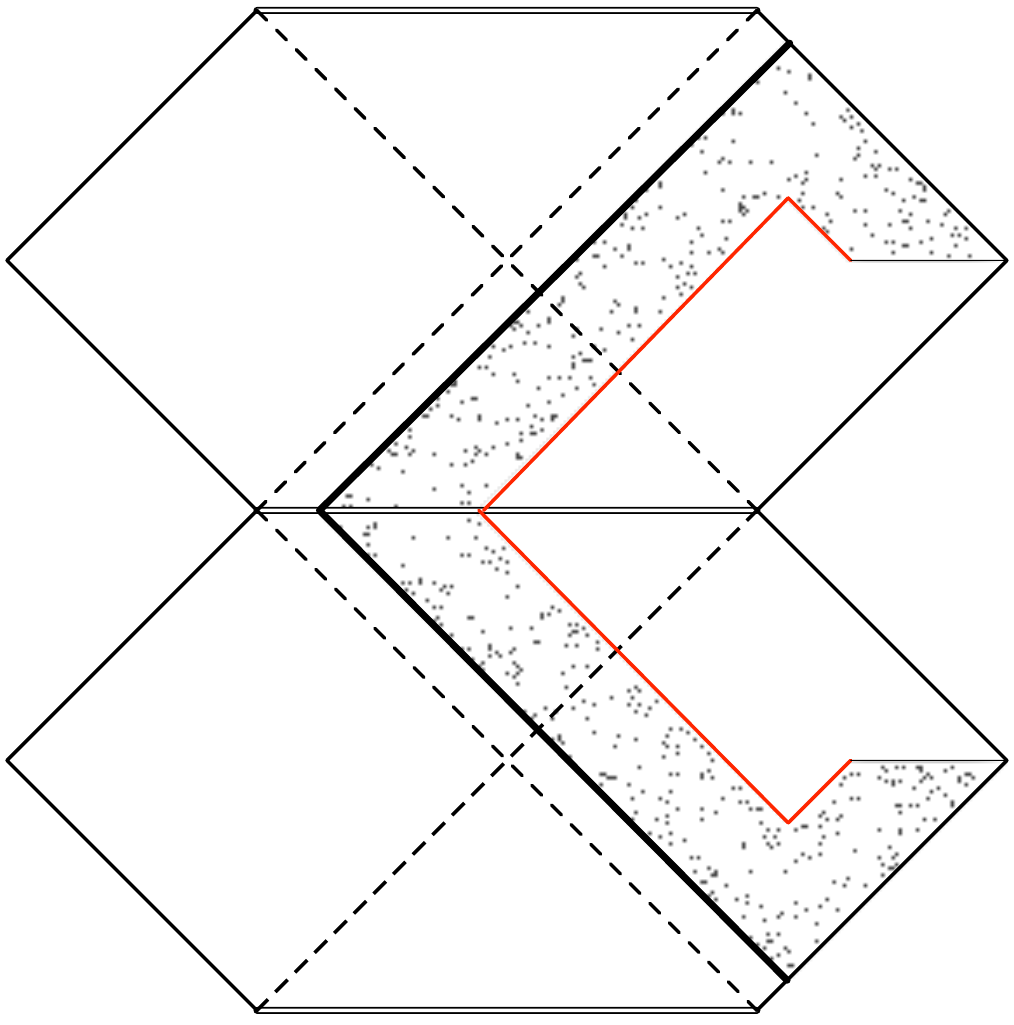}\hspace{1.6cm} \includegraphics[height=3cm]{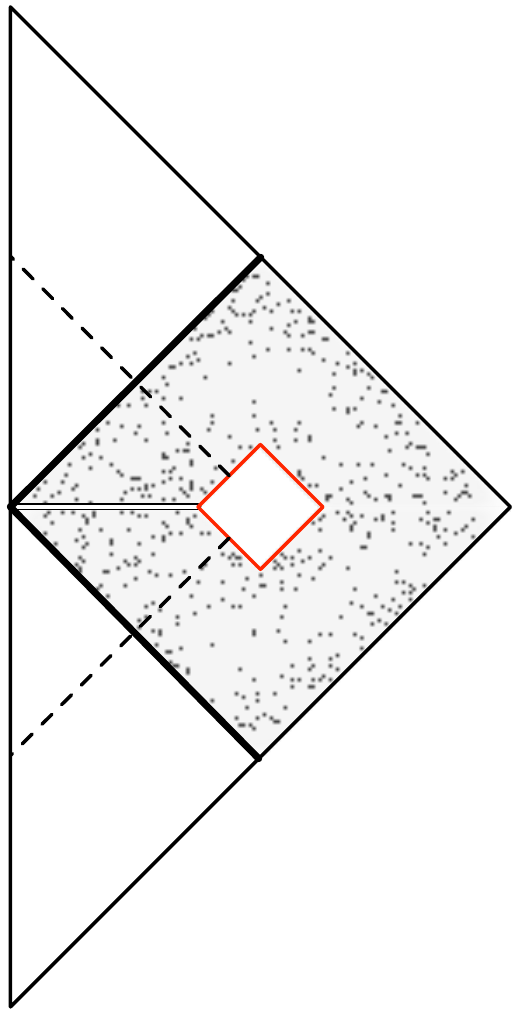}  
\vspace{-1em}
\caption{\em Left: Two Kruskal spacetimes are glued at the singularity. The grey region is the metric of a black to white hole transition outside a collapsing and the exploding null shell. Right: The corresponding regions in the physical spacetime.}
\label{12}
\end{figure}

Disregarding Hawking evaporation, a metric describing this process outside the $B$ region can be written explicitly by cutting and pasting the extended Schwarzschild solution, following \cite{Haggard2014}.  This is illustrated in Figure \ref{12}: two Kruskal spacetimes are glued across the singularity as described in the previous section and the shaded region is the metric of the portion of spacetime outside a collapsing shell (here chosen to be null).

While the location of the $A$ region is determined by the classical theory, the location of the $B$ region, instead, is determined by quantum theory. The $B$ process is indeed a typical quantum tunneling process: it has a long lifetime.   A priori, the value of $\tau_{bh}$ is determined probabilistically by quantum theory.  As in conventional tunneling, in a stationary situation (when the horizon area varies slowly), we expect the probability $p$ per unit time for the tunneling to happen to be time independent.  This implies that the normalised probability $P(t)$ that the tunneling happens between times $t$ and $t+dt$ is governed by $dP(t)/dt=-p P(t)$, namely is 
\be
P(t)=\frac{1}{\tau_{bh}}e^{-\frac{t}{\tau_{bh}}},
\ee
which is normalised ($\int_0^\infty P(t) dt =1$) and where $\tau_{bh}$ satisfies 
\be
\tau_{bh}=1/p. 
\ee

We note parenthetically that the quantum spread in the lifetime can be a source of apparent unitarity violation, for the following reason. In conventional nuclear decay, a tunneling phenomenon, the quantum indetermination in the decay time is of the same order as the lifetime.  The unitary evolution of the state of a particle trapped in the nucleus is such that the state slowly leaks out, spreading it over a vast region. A Geiger counter has a small probability of detecting a particle at the time where it happens to be. Once the detection happens, there is an apparent violation of unitarity. (In the Copenhagen language the Geiger counter measures the state, causing it to collapse, loosing information. In the Many Worlds language, the state splits into a continuum of branches that decohere and the information of a single branch is less than the initial total information.) In either case, the evolution of the quantum state from the nucleus to a \emph{given} Geiger counter detection is not unitary; unitarity is recovered by taking into account the full spread of different detection times.  The same must be true for the tunneling that disrupts the black hole. If tunneling will happen at a time $t$, unitarity can only be recovered by taking into account the full quantum spread of the tunneling time, which is to say: over different future goemetries.  The quantum state is actually given by a quantum superposition of a continuum of spacetimes as in Figure \ref{9}, each with a different value of $v_-$ and $v_+$.  We shall not further pursue here the analysis of this apparent source of unitarity, but we indicate it for future reference. 

\section{The $B$ region: the Horizon at the Transition} 

\begin{figure}[h]
\includegraphics[height=3.25cm]{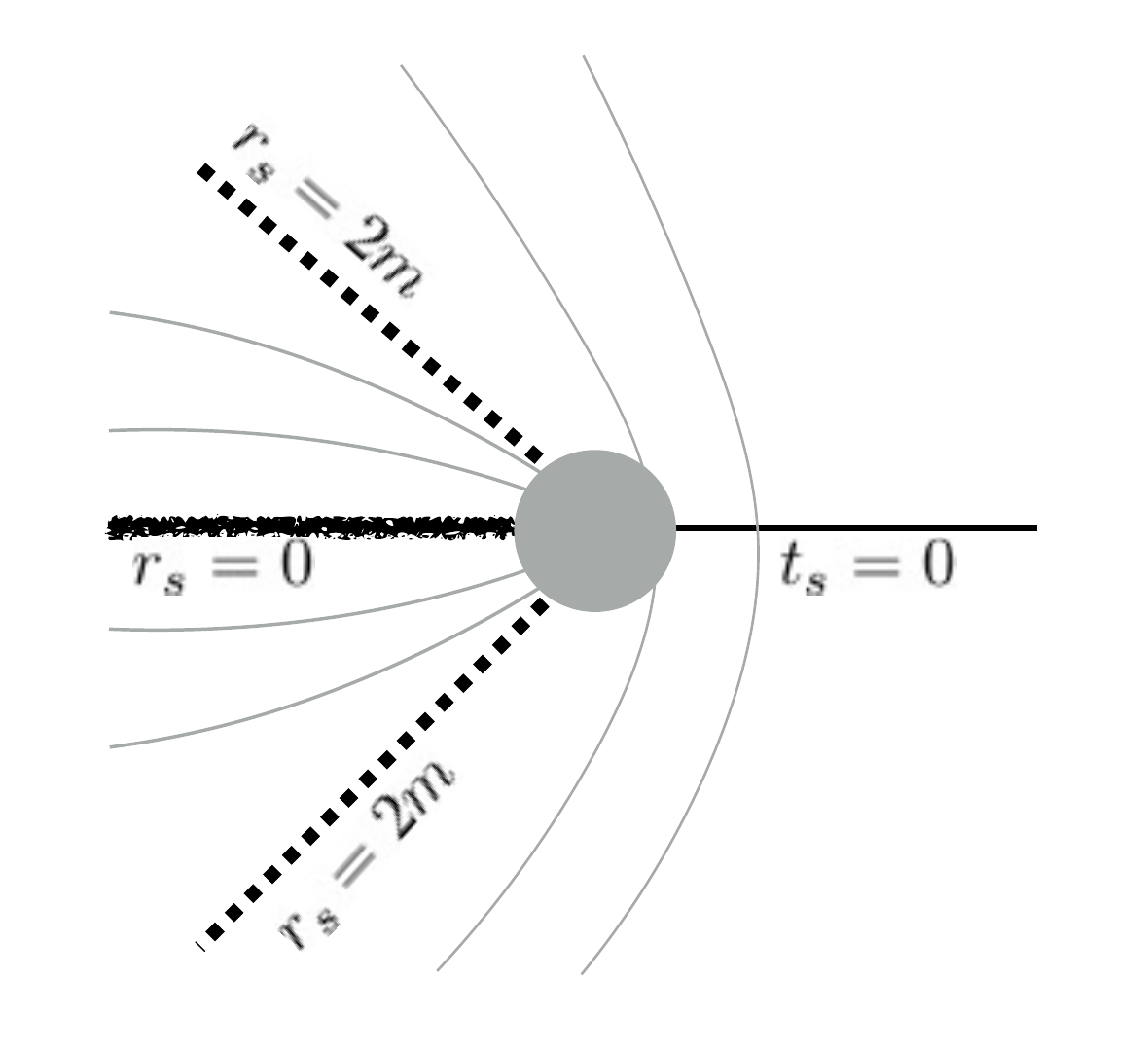} \hspace{.5cm} \includegraphics[height=3cm]{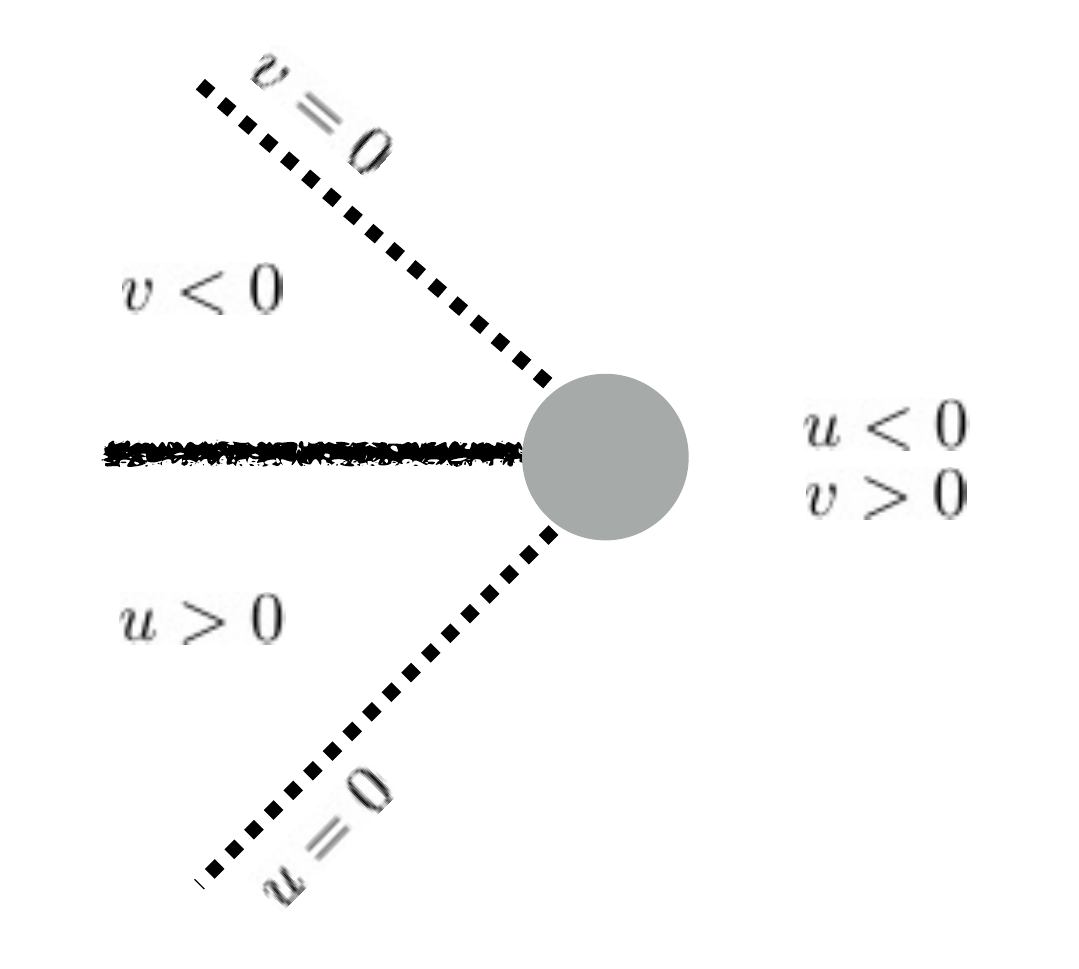}
\vspace{-1em}
\caption{\em The $B$ region. Left: Surfaces of equal Schwarzschild radius are depicted. Right: The signs of the null Kruskal coordinates around $B$.}
\label{11}
\end{figure}

The geometry surrounding the transition in the $B$ region is depicted in detail in Figure \ref{11}.  
The metric of the entire neighbourhood of the $B$ region is an extended Schwarzschild metric. It can therefore be written in null Kruskal coordinates 
\be
ds^2=-\frac{32m^3}{r}e^{-\frac{r}{2m}} du dv + r^2 d\Omega^2,\label{kr}
\ee
where 
\be
 \left(1-\frac{r}{2m}\right)e^{\frac{r}{2m}}=uv.
 \label{r}
\ee
On the two horizons we have respectively $v=0$ and $u=0$, and separate regions where $u$ and $v$ have different signs as in the right panel of Figure \ref{11}.  Notice the rapid change of the value of the radius across the $B$ region, which yields a rapid variation of the metric components in \eqref{kr}. 

\begin{figure}[h]
\includegraphics[height=4cm]{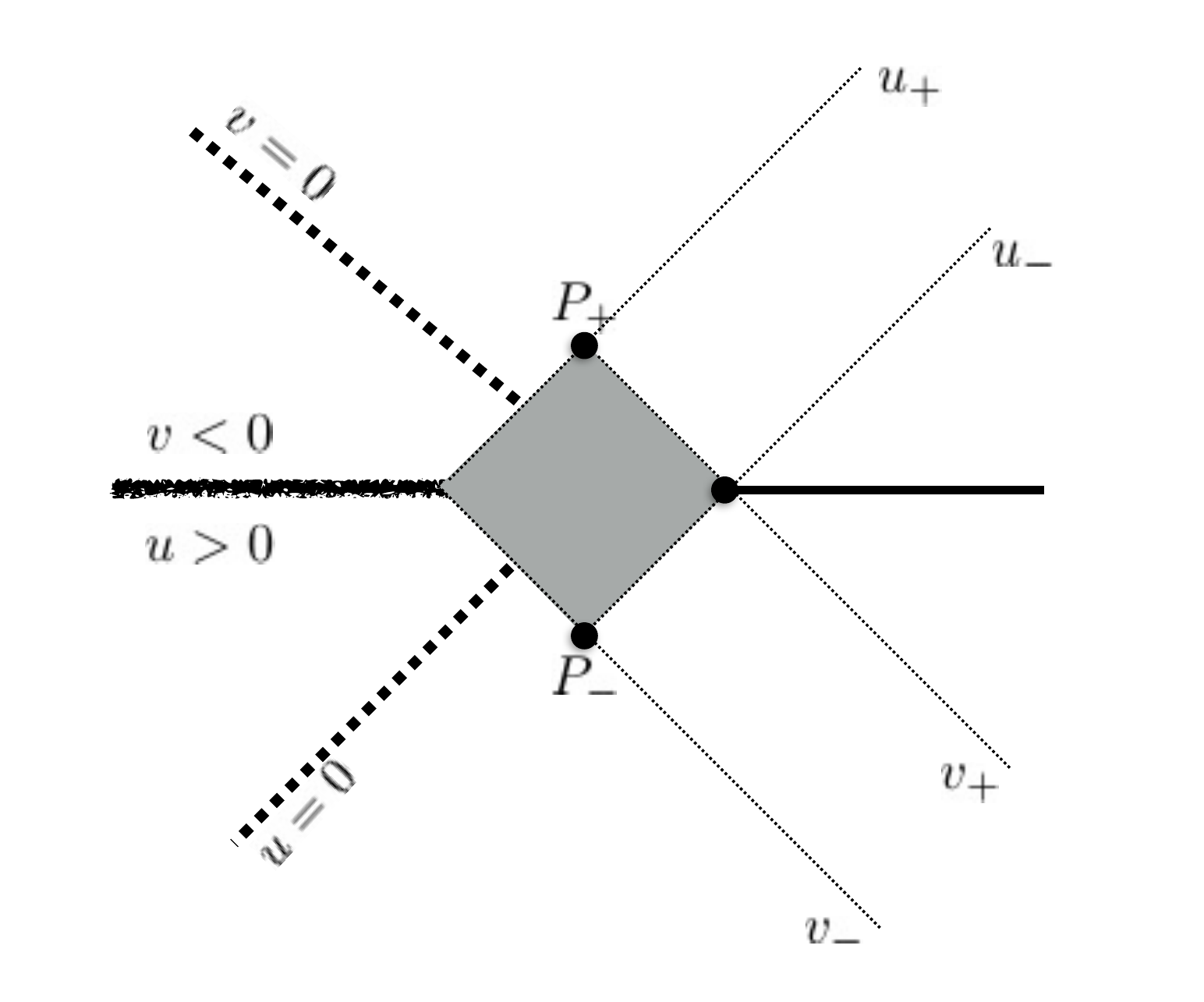}
\vspace{-2em}
\caption{\em The $B$ transition region.}
\label{PPP}
\end{figure}

To fix the region $B$, we need to specify more precisely its boundary, which we have not done so far. It is possible to do so by identifying it with the diamond (in the 2d diagram) defined by two points $P_+$ and $P_{-}$ with coordinates $v_\pm, u_\pm$ both outside the horizon, at the same radius $r_P$, and at opposite timelike distance from the bounce time, see Figure \ref{PPP}. 

The same radius $r_P$ implies 
\be
v_+u_+=v_-u_-\equiv \left(1-\frac{r_P}{2m}\right)e^{\frac{r_P}{2m}}.
\ee
The same time from the horizon implies that the light lines $u=u_-$ and $v=v_+$ cross on $t_s=0$, or $u+v=0$, hence
\be
u_-=-v_+.
\ee
This crossing point is the outermost reach of the quantum region, with radius $r_m$ determined by
\be
v_+u_- \equiv \left(1-\frac{r_m}{2m}\right)e^{\frac{r_m}{2m}}. 
\ee
The region is then entirely specified by two parameters. We can take them to be $r_P$ and $\Delta\tau = v_+-v_-\sim u_+-u_-$. The first characterizes the radius at which the  quantum transition starts. The second its duration. (Strictly speaking, we could also have $ v_+-v_-$ and $u_+-u_-$ of different orders of magnitude, but we do not explore this possibility here.)  

There are indications about both metric scales in the literature. In \cite{Haggard2014,Haggard2016}, arguments where given for $r_P\sim 7/3 \ m$.  Following \cite{Christodoulou2016}, the duration of the transition has been called ``crossing time" and computed by Christodoulou and D'Ambrosio in \cite{Christodoulou2018,Marios:2018} using Loop Quantum Gravity: the result is $\Delta \tau \sim m$, which can be taken as a confirmation of earlier results \cite{Ambrus2005,Barcelo2014b,Barcelo2016} obtained with other methods. The two crucial remaining parameters are the black hole and the white hole lifetimes, $\tau_{bh}$ and $\tau_{wh}$.

The result in \cite{Christodoulou2018} indicates also that $p$, the probability of tunneling per unit time,  is suppressed exponentially by a factor $e^{-m^2/\hbar}$. Here  $m$ is not the initial mass $m_o$ of the black hole at the time of its formation, rather, it is the mass of the black hole at the decay time.  
This is in accord with the semiclassical estimate that tunneling is suppressed as in (\ref{suppression}) and (\ref{suppression2}).  As mentioned in the introduction, because of Hawking evaporation, the mass of the black hole shrinks to Planckian values in a time of order $m_o^3/\hbar$, where the probability density becomes of order unit, giving
\be
\tau_{{bh}}\sim m_o^3/\hbar
\ee
and 
\be
\Delta\tau\sim \sqrt{\hbar}.
\ee 
We conclude that region $B$ has a Planckian size.
 
We notice parenthetically that the value of $p$ above is at odds with the arguments given in \cite{Haggard2014} for a shorter lifetime $\tau_{{bh}}\sim m_o^2/\sqrt{\hbar}$. This might be because the analysis in \cite{Christodoulou2018} captures the dynamics of only a few of the relevant degrees of freedom, but we do not consider this possibility here. The entire range of possibilities for the black to white transition lifetime, $
m_o^2/\sqrt{\hbar}\le\tau_{{bh}}\le m_o^3/\hbar$, may have phenomenological consequences, which have been explored in \cite{Barrau2014c,Barrau2014b,Barrau2015,Barrau2016,Rovelli2017}. (On hypothetical white hole observations see also \cite{Retter2012}).  

\section{Interior Volume and Purification Time} 

Consider a quantum field living on the background geometry described above.  Near the black hole horizon there is production of Hawking radiation.  Its back-reaction on the geometry gradually decreases the area of the horizon.  This, in turn, increases the transition probability to a white hole. After a time $\tau_{{bh}}\sim m_o^3/\hbar$, the area of the black hole reaches the Planckian scale $A_{{bh}}(\text{final})\sim \hbar$, and the transition probability becomes of order unity.  The volume of the transition surface is huge. 

To compute it with precision, we should compute the back-reaction of the inside component of the Hawking radiation, which gradually decreases the value of $m$ as the coordinate $x$ increases.  Intuitively, the inside components of the Hawking pairs fall toward the singularity, decreasing $m$.  Since most of the decrease is at the end of the process, we may approximate the full interior of the hole with that of a Schwarzschild solution of mass $m_o$ and the first order estimate of the inside volume should not be affected by this process. Thus we may assume that the volume at the transition has the same order as the one derived in Eq. \eqref{V}, namely 
\be
V_{{bh}}(\text{final})\sim \sqrt{\hbar} \, m_o \, \tau_{bh} \sim m_o^4/\sqrt{\hbar}.\label{finalV}
\ee   
Using the same logic in the future of the transition, we approximate the inside metric of the white hole with that of a Schwarzschild solution of Planckian mass, since in the future of the singularity, the metric is again of Kruskal type, but now for a white hole of Plankian mass.   

The last parameter to estimate is the lifetime $\tau_{{wh}}=u_0-u_+$ of the white hole produced by the transition. To do so, we can assume that the internal volume is conserved in the quantum transition.  The volume of the region of Planckian curvature inside the white hole horizon is then
\begin{equation}
V_{{wh}}(u)\sim l^2\sqrt{\frac{m}{l}}\; \tau_{wh}, \label{Vw}
\end{equation}
where now  $l\sim m \sim \sqrt{\hbar},$ and therefore
\be
V_{wh}(\text{initial}) \sim \hbar \, \tau_{wh}.
\ee
Gluing the geometry on the past side of the singularity to the geometry on the future side requires that the two volumes match, namely that \eqref{Vw} matches \eqref{V} and this gives 
\begin{equation}
\tau_{{wh}}\sim m_o^4/\hbar^{3/2}. 
\label{tauwh}
\end{equation}
This shows that the Planck-mass white hole is a long-lived remnant \cite{Christodoulou2016a}. 

With these results, we can address the black hole information paradox. The Hawking radiation reaches future infinity before $u_-$, and is described by a \emph{mixed} state with an entropy of order $m_o^2/\hbar$. This must be purified by correlations with field excitations inside the hole.  In spite of the smallness of the mass of the hole, the large internal volume \eqref{finalV} is sufficient to host these excitations \cite{Rovelli2017a}.  This addresses the requirement (a) of the introduction, namely that there is a large information capacity.   

To release this entropy, the remnant must be long-lived. During this time, any internal information that was trapped by the black hole horizon can leak out.  Intuitively, the interior member of a Hawking pair can now escape and purify the exterior quantum state.  The long lifetime of the white hole allows this information to escape in the form of very low frequency particles, thus respecting bounds on the maximal entropy contained in a given volume with given energy.   

The lower bound imposed by unitarity and energy considerations is $\tau_R\sim m_o^4/\hbar^{3/2}$ \cite{Preskill1993,Bianchi:2014bma,Marolf2017} and this is precisely the white hole lifetime \eqref{tauwh} deduced above; hence we see that they satisfy the requirement (b) of the introduction. Therefore white holes realize precisely the long-lived remnant scenario for the end of the black hole evaporation that was conjectured and discussed mostly in the 1990's \cite{Banks1992,Giddings1992a,Giddings1992c,Giddings1993,Giddings1994,Banks1993,Banks1993b,Banks1995}. 

The last issue we should discuss is stability.  Generically, white holes are known to be unstable under perturbations  (see for instance Chapter 15 in \cite{Frolov2012} and references therein). The instability arises because modes of short-wavelength are exponentially blue-shifted along the white hole horizon. In the present case, however, we have a Planck-size white hole.  To run this argument for instability in the case of a planckian white hole, it is necessary to consider transplanckian perturbations. Assuming no transplanckian perturbations to exist, there are no instabilities to be considered. This addresses the requirement (c). Alternatively: a white hole is unstable because it may re-collapse into a black hole with similar mass; therefore a Planck size white hole can at most re-collapse into a Planck size black hole; but this has probability of order unity to tunnel back into a white hole in a Planck time.

Therefore the proposed scenario addresses the consistency requirements (a), (b), and (c) for the solution of the information-loss paradox and provides an effective geometry for the end-point of black hole evaporation: a long-lived Planck-mass white hole.

\section{On White Holes}

Notice that from the outside, a white hole is indistinguishable from a black hole. This is obvious from the existence of the Kruskal spacetime, where the same region of spacetime (region I) describes both the exterior of a black hole and the exterior of a white hole.  For $r_s\!>\!2m$, the conventional Schwarzschild line element describes equally well a black hole exterior and a white hole exterior. The difference is only what happens at $r=2m$. 

The only locally salient difference between a white and a black hole is that if we add some \emph{generic} perturbation or matter on a given constant $t_s$ surface, in (the Schwarzschild coordinate description of) a black hole we see matter falling towards the center and accumulating around the horizon.  While in (the Schwarzschild coordinate description of) a white hole we see matter accumulated around the horizon in the past, moving away from the center. Therefore the distinction is only one of ``naturalness" of initial conditions: a black hole has ``special" boundary conditions in the future, a white hole has ``special" boundary conditions in the past. 

This difference can be described  physically also as follows: if we look at a black hole (for instance when the Event Horizon Telescope \cite{Doeleman2009} examines Sagittarius A*), we see a black disk.  This means that generic initial conditions on past null infinity give rise on future null infinity to a black spot with minimal incoming radiation: a ``special" configuration in the future sky.   By time reversal symmetry, the opposite is true for a white hole; generic initial conditions on future null infinity require a black spot with minimal incoming radiation from past null infinity: a ``special" configuration in the past.   

We close this section by briefly discussing the ``no transition principle" considered by Engelhardt and Horowitz in \cite{Engelhardt2016}. By assuming ``holographic" unitarity at infinity and observing that consequently information cannot leak out from the spacetime enclosed by a single asymptotic region, these authors rule out a number of potential scenarios, including the possibility of resolving generic singularities inside black holes.  Remarkably, the scenario described here circumvents the no transition principle  and permits singularity resolution in the bulk: the reason is that this singularity is confined in a finite spacetime region and does not alter the global causal structure. 

\section{On remnants}

The long-lived remnant scenario provides a satisfactory solution to the black-hole information paradox.  The main reason for which it was largely discarded was the fact that remnants appeared to be exotic objects extraneous to known physics.  Here we have shown that they are not: white holes are well known solutions of the Einstein equations and they provide a concrete model for long-lived remnants. 

Two other arguments made long-lived remnants unpopular: Page's version of the information paradox; and the fact that if remnants existed they would easily be produced in accelerators.   Neither of these arguments applies to the long-lived remnant scenario of this paper.  We discuss them below. 

In its interactions with its surroundings, a black hole with horizon area $A$ behaves thermally as a system with entropy $S_{bh}=A/4\hbar$. This is a fact supported by a large number of convincing arguments and continues to hold for the dynamical horizons we consider here. The Bekenstein-Hawking entropy provides a good notion of entropy that satisfies Bekenstein's generalized second law, {\em in the approximation in which we can treat the horizon as an event horizon}.  In the white hole remnant scenario this is a good approximation for a long time, but fails at the Planck scale when the black hole transitions to a white hole. 

  Let us assume for the moment that these facts imply the following hypothesis (see for instance \cite{Marolf2017})  
\begin{quote} 
(H) The total number of available states for a quantum  system living on the internal spatial slice $\Sigma_i$ of Figure \ref{uno}  is $N_{bh}=e^{S_{bh}}=e^{A/4\hbar}$.  
\end{quote}
Then, as noticed by Page \cite{Page1993a}, we have immediately an information paradox regardless of what happens at the end of the evaporation. The reason is that the entropy of the Hawking radiation grows with time.  It is natural to interpret this entropy as correlation entropy with the Hawking quanta that have fallen inside the hole, but for this to happen there must be a sufficient number of available states inside the hole.  If hypothesis (H) above is true, then this cannot be, because as the area of the horizon decreases with time, the number of available internal states decreases and becomes insufficient to purify the Hawking radiation.  The time at which the entropy surpasses the area is known as the Page time.  This has lead many to hypothesize that the Hawking radiation is already purifying itself by the Page time: a consequence of this idea is the firewall scenario \cite{Almheiri2013}. 

The hypothesis (H) does not apply to the white-hole remnants. As argued in \cite{Rovelli2017a}, growing interior volumes together with the existence of local observables implies that the number of internal states grows with time instead of decreasing as stated in (H).  This is not in contradiction with the fact that a black hole behaves thermally in its interactions with its surroundings as a system with entropy $S=A/4\hbar$.  The reason is that ``entropy" is not an absolute concept and the notion of entropy must be qualified.   Any definition of ``entropy" relies on a coarse graining, namely on ignoring some variables: these could be microscopic variables, as in the statistical mechanical notion of entropy, or the variables of a subsystem over which we trace, as in the von Neumann entropy.  The Bekenstein-Hawking entropy correctly describes the thermal interactions of the hole with its surroundings, because the boundary is an outgoing null surface and $S_{bh}$ counts the number of states {\em that can be distinguished from the exterior}; but this is not the number of states that can be distinguished by local quantum field operators on $\Sigma_i$  \cite{Rovelli2017a}. See also \cite{Giddings2013}.

Therefore there is no reason for the Hawking radiation to purify itself by the Page time. This point has been stressed by Unruh and Wald in their discussion of the evaporation process on the spacetime pictured in the left panel of Figure \ref{7}, see e.g. \cite{Unruh2017}. Our scenario differs from Unruh and Wald's  in that the white hole transition allows the Hawking partners that fell into the black hole to emerge later and purify the state. They emerge slowly, over a time of order $m_o^4/\hbar^{3/2}$, in a manner consistent with the long life of the white hole established here. 

The second standard argument against remnants is that, if they existed, it would be easy to produce them.  This argument assumes that a remnant has a small boundary area and little energy, but can have a very large number of states.  The large number of states would contribute a large phase-space volume factor in any scattering process, making the production of these objects in scattering processes highly probable.  Actually, since in principle these remnants could have an \emph{arbitrarily} large number of states, their phase-space volume factor would be infinite, and hence they would be produced spontaneously everywhere. 

This argument does not apply to white holes. The reason is that a white hole is screened by an anti-trapping horizon: the only way to produce it is through quantum gravity tunneling from a black hole!   Even more, to produce a Planck mass white hole with a large interior volume, we must first produce a \emph{large} black hole and let it evaporate for a long time.  Therefore the threshold to access the full phase-space volume of white holes is high.  A related argument is in \cite{Banks1993}, based on the fact that infinite production rate is prevented by locality.   In \cite{Giddings1994} Giddings questions this point treating remnants as particles of an effective field theory; the field theory, however, may be a good approximation of such a highly non-local structure as a large white hole only in the approximation where the large number of internal states is not seen.  See also \cite{Banks1995}.\\

\section{Conclusion}

As a black hole evaporates, the probability to tunnel into a white hole increases.  The suppression factor for this tunneling process is of order $e^{-{m^2}/{m^2_{Pl}}}$.  Before reaching sub-Planckian size, the probability ceases to be suppressed and the black hole tunnels into a white hole.

Old black holes have a large volume. Quantum gravitational tunneling results in a Planck-mass white hole that also has a large interior volume. The white hole is long-lived because it takes awhile for its finite, but large, interior to become visible from infinity. 

The geometry outside the black to white hole transition is described by a single asymptotically-flat spacetime. The Einstein equations are violated in two regions: The Planck-curvature region A, for which we have given an effective metric that smoothes out of the singularity; and the tunneling region B, whose size and decay probability can be computed \cite{Christodoulou2018}.  These ingredients combine to give a white hole remnant scenario. 

This scenario provides a way to address the information problem. We distinguish two ways of encoding information, the first associated with the small area of the horizon and the second associated to the remnant's interior.  The Bekenstein-Hawking entropy $S_{bh}=A/4\hbar$ is encoded on the horizon and counts states that can only be distinguished from outside. On the other hand, a white hole resulting from a quantum gravity transition has a large volume that is available to encode substantial information even when the horizon area is small. The white hole scenario's apparent horizon, in contrast to an event horizon, allows for information to be released. The long-lived white hole releases this information slowly and purifies the Hawking radiation emitted during evaporation. Quantum gravity resolves the information problem. 
  
 \centerline{---}

CR thanks Ted Jacobson, Steve Giddings, Gary Horowitz, Steve Carlip, and Claus Kiefer for very useful exchanges during the preparation of this work. EB and HMH thank Tommaso De Lorenzo for discussion of time scales.  EB thanks Abhay Ashtekar for discussion of remnants. HMH thanks the CPT for warm hospitality and support, Bard College for extended support to visit the CPT with students, and the Perimeter Institute for Theoretical Physics for generous sabbatical support. MC acknowledges support from the SM Center for Space, Time and the Quantum and the Leventis Educational Grants Scheme. This work is  supported  by  Perimeter  Institute  for  Theoretical  Physics.   Research  at  Perimeter  Institute is supported by the Government of Canada through Industry Canada and by the Province of Ontario through the Ministry of Research and Innovation.

\bibliographystyle{utcaps}

\bibliography{/Users/carlorovelli/Documents/library}

\providecommand{\href}[2]{#2}\begingroup\raggedright\begin{thebibliography}{10}

\bibitem{Hawking1974}
S.~W. Hawking, ``{Black hole explosions?},''
  \href{http://dx.doi.org/10.1038/248030a0}{{\em Nature} {\bf 248} (1974)
  30--31}.

\bibitem{Rovelli2014}
C.~Rovelli and F.~Vidotto, ``{Planck stars},'' {\em Int. J. Mod. Phys. D} {\bf
  23} (2014)  1442026, \href{http://arxiv.org/abs/1401.6562}{{\tt
  arXiv:1401.6562}}.

\bibitem{Haggard2014}
H.~M. Haggard and C.~Rovelli, ``{Black hole fireworks: quantum-gravity effects
  outside the horizon spark black to white hole tunneling},'' {\em Physical
  Review} {\bf D92} (2015)  104020, \href{http://arxiv.org/abs/1407.0989}{{\tt
  arXiv:1407.0989}}.

\bibitem{DeLorenzo2016}
T.~{De Lorenzo} and A.~Perez, ``{Improved black hole fireworks: Asymmetric
  black-hole-to-white-hole tunneling scenario},'' {\em Physical Review D} {\bf
  93} (2016)  124018, \href{http://arxiv.org/abs/1512.04566}{{\tt
  arXiv:1512.04566}}.

\bibitem{Christodoulou2016}
M.~Christodoulou, C.~Rovelli, S.~Speziale, and I.~Vilensky, ``{Planck star
  tunneling time: An astrophysically relevant observable from background-free
  quantum gravity},'' {\em Physical Review D} {\bf 94} (2016)  084035,
  \href{http://arxiv.org/abs/1605.05268}{{\tt arXiv:1605.05268}}.

\bibitem{Christodoulou2018}
M.~Christodoulou and F.~D'Ambrosio, ``{Characteristic Time Scales for the
  Geometry Transition of a Black Hole to a White Hole from Spinfoams},''
  \href{http://arxiv.org/abs/1801.03027}{{\tt arXiv:1801.03027}}.
  
\bibitem{Marios:2018}
M. Christodoulou,
``{Geometry Transition in Covariant Loop Quantum Gravity},"
\href{http://arxiv.org/abs/1803.00332}{{\tt arXiv:1803.00332 [gr-qc]}}

\bibitem{Frolov:1979tu}
V.~P. Frolov and G.~Vilkovisky, ``{Quantum Gravity removes Classical
  Singularities and Shortens the Life of Black Holes},'' {\em ICTP preprint
  IC/79/69, Trieste.} (1979)  .

\bibitem{Frolov:1981mz}
V.~Frolov and G.~Vilkovisky, ``{Spherically symmetric collapse in quantum
  gravity},'' {\em Physics Letters B} {\bf 106} (1981)  307--313.

\bibitem{Stephens1994}
C.~R. Stephens, G.~t. Hooft, and B.~F. Whiting, ``{Black hole evaporation
  without information loss},'' {\em Classical and Quantum Gravity} {\bf 11}
  (1994)  621--647, \href{http://arxiv.org/abs/9310006}{{\tt arXiv:9310006
  [gr-qc]}}.

\bibitem{modesto2004disappearance}
L.~Modesto, ``{Disappearance of the black hole singularity in loop quantum
  gravity},'' {\em Physical Review D} {\bf 70} (2004) no.~12, 124009.

\bibitem{Modesto2006a}
L.~Modesto, ``{Evaporating loop quantum black hole},''
  \href{http://arxiv.org/abs/0612084}{{\tt arXiv:0612084 [gr-qc]}}.

\bibitem{Mazur:2004}
P.~O. Mazur and E.~Mottola, ``{Gravitational vacuum condensate stars.},'' {\em
  Proceedings of the National Academy of Sciences of the United States of
  America} {\bf 101} (2004) no.~26, 9545--50,
  \href{http://arxiv.org/abs/0407075}{{\tt arXiv:0407075 [gr-qc]}}.

\bibitem{Ashtekar:2005cj}
A.~Ashtekar and M.~Bojowald, ``{Black hole evaporation: A paradigm},'' {\em
  Class. Quant. Grav.} {\bf 22} (2005)  3349--3362,
  \href{http://arxiv.org/abs/0504029}{{\tt arXiv:0504029 [gr-qc]}}.

\bibitem{Balasubramanian:2006}
V.~Balasubramanian, D.~Marolf, and {Rozali. M.}, ``{Information Recovery From
  Black Holes},'' {\em Gen. Rel. Grav.} {\bf 38} (2006)  1529--1536,
  \href{http://arxiv.org/abs/0604045 [gr-qc]}{{\tt arXiv:0604045 [gr-qc]}}.

\bibitem{Hayward2006}
S.~A. Hayward, ``{Formation and Evaporation of Nonsingular Black Holes},'' {\em
  Phys. Rev. Lett.} {\bf 96} (2006)  031103,
  \href{http://arxiv.org/abs/0506126}{{\tt arXiv:0506126 [gr-qc]}}.

\bibitem{Hossenfelder:2009fc}
S.~Hossenfelder, L.~Modesto, and I.~Premont-Schwarz, ``{A model for
  non-singular black hole collapse and evaporation},'' {\em Phys. Rev.} {\bf
  D81} (2010)  44036, \href{http://arxiv.org/abs/0912.1823}{{\tt
  arXiv:0912.1823}}.

\bibitem{Hossenfelder:2010a}
S.~Hossenfelder and L.~Smolin, ``{Conservative solutions to the black hole
  information problem},'' {\em Physical Review D} {\bf 81} (2010)  064009,
  \href{http://arxiv.org/abs/0901.3156}{{\tt arXiv:0901.3156}}.

\bibitem{frolov:BHclosed}
V.~P. Frolov, ``{Information loss problem and a ``black hole'' model with a
  closed apparent horizon},'' \href{http://arxiv.org/abs/1402.5446}{{\tt
  arXiv:1402.5446}}.

\bibitem{GambiniPullin2014a}
R.~Gambini and J.~Pullin, ``{A scenario for black hole evaporation on a quantum
  Geometry},'' {\em Proceedings of Science} {\bf 15-18-July} (2014)  ,
  \href{http://arxiv.org/abs/1408.3050}{{\tt arXiv:1408.3050}}.

\bibitem{GambiniPullin2014b}
R.~Gambini and J.~Pullin, ``{Quantum shells in a quantum space-time},''
  \href{http://dx.doi.org/10.1088/0264-9381/32/3/035003}{{\em Classical and
  Quantum Gravity} {\bf 32} (2015) no.~3, },
  \href{http://arxiv.org/abs/1408.4635}{{\tt arXiv:1408.4635}}.

\bibitem{Bardeen2014}
J.~M. Bardeen, ``{Black hole evaporation without an event horizon},''
  \href{http://arxiv.org/abs/1406.4098}{{\tt arXiv:1406.4098}}.

\bibitem{Giddings1992b}
S.~B. Giddings and W.~M. Nelson, ``{Quantum emission from two-dimensional black
  holes},'' {\em Physical Review D} {\bf 46} (1992)  2486--2496,
  \href{http://arxiv.org/abs/9204072}{{\tt arXiv:9204072 [hep-th]}}.

\bibitem{Narlikar1974}
J.~V. Narlikar, K.~{Appa Rao}, and N.~Dadhich, ``{High energy radiation from
  white holes},'' {\em Nature} {\bf 251} (1974)  591.

\bibitem{HAJICEK2001}
P.~H{\'{a}}j{\'{i}}{\v{c}}ek and C.~Kiefer, ``{Singularity avoidance by
  collapsing shells in quantum gravity},'' {\em International Journal of Modern
  Physics D} {\bf 10} (2001) no.~06, 775--779,
  \href{http://arxiv.org/abs/0107102}{{\tt arXiv:0107102 [gr-qc]}}.

\bibitem{Ambrus2005}
M.~Ambrus and P.~H{\'{a}}j{\'{i}}{\v{c}}ek, ``{Quantum superposition principle
  and gravitational collapse: Scattering times for spherical shells},'' {\em
  Physical Review D} {\bf 72} (2005)  064025,
  \href{http://arxiv.org/abs/0507017}{{\tt arXiv:0507017 [gr-qc]}}.

\bibitem{Olmedo:2017lvt}
J.~Olmedo, S.~Saini, and P.~Singh, ``{From black holes to white holes: a
  quantum gravitational, symmetric bounce},''
  \href{http://dx.doi.org/10.1088/1361-6382/aa8da8}{{\em Class. Quant. Grav.}
  {\bf 34} (2017) no.~22, 225011}, \href{http://arxiv.org/abs/1707.07333}{{\tt
  arXiv:1707.07333 [gr-qc]}}.

\bibitem{Aharonov1987}
Y.~Aharonov, A.~Casher, and S.~Nussinov, ``{The unitarity puzzle and Planck
  mass stable particles},'' {\em Physics Letters B} {\bf 191} (1987) no.~1-2,
  51--55.
  \url{https://www.sciencedirect.com/science/article/pii/0370269387913207}.

\bibitem{Giddings1992a}
S.~Giddings, ``{Black holes and massive remnants},'' {\em Physical Review D}
  {\bf 46} (1992) no.~4, 1347--1352, \href{http://arxiv.org/abs/9203059}{{\tt
  arXiv:9203059 [hep-th]}}.

\bibitem{Callan1992}
C.~G. Callan, S.~B. Giddings, J.~A. Harvey, and A.~Strominger, ``{Evanescent
  black holes},'' \href{http://dx.doi.org/10.1103/PhysRevD.45.R1005}{{\em
  Physical Review D} {\bf 45} (1992) no.~4, R1005--R1009}.
  \url{https://link.aps.org/doi/10.1103/PhysRevD.45.R1005}.

\bibitem{Giddings1993}
S.~B. Giddings, ``{Constraints on black hole remnants},'' {\em Physical Review
  D} {\bf 49} (1994)  947--957, \href{http://arxiv.org/abs/9304027}{{\tt
  arXiv:9304027 [hep-th]}}.

\bibitem{Preskill1993}
J.~Preskill, ``{Do Black Holes Destroy Information?},'' in {\em An
  international symposium on Black Holes, Membranes, Wormholes and
  Superstrings}, S.~Kalara and D.~Nanopoulos, eds., p.~22.
\newblock World Scientific, Singapore, 1993.

\bibitem{Banks1993}
T.~Banks and M.~O'Loughlin, ``{Classical and quantum production of cornucopions
  at energies below 1018 GeV},'' {\em Physical Review D} {\bf 47} (1993)
  540--553, \href{http://arxiv.org/abs/9206055}{{\tt arXiv:9206055 [hep-th]}}.

\bibitem{Banks1995}
T.~Banks, ``{Lectures on black holes and information loss},'' {\em Nuclear
  Physics B (Proceedings Supplements)} {\bf 41} (1995) no.~1-3, 21--65,
  \href{http://arxiv.org/abs/9412131}{{\tt arXiv:9412131 [hep-th]}}.

\bibitem{Ashtekar:2008jd}
A.~Ashtekar, V.~Taveras, and M.~Varadarajan, ``{Information is Not Lost in the
  Evaporation of 2-dimensional Black Holes},'' {\em Phys.Rev.Lett.} {\bf 100}
  (2008)  211302.

\bibitem{Ashtekar:2010qz}
A.~Ashtekar, F.~Pretorius, and F.~M. Ramazanoglu, ``{Evaporation of
  2-Dimensional Black Holes},''
  \href{http://dx.doi.org/10.1103/PhysRevD.83.044040}{{\em Phys. Rev.} {\bf
  D83} (2011)  44040}, \href{http://arxiv.org/abs/1012.0077}{{\tt
  arXiv:1012.0077 [gr-qc]}}.

\bibitem{Ashtekar:2010hx}
A.~Ashtekar, F.~Pretorius, and F.~M. Ramazanoglu, ``{Surprises in the
  Evaporation of 2-Dimensional Black Holes},''
  \href{http://dx.doi.org/10.1103/PhysRevLett.106.161303}{{\em Phys. Rev.
  Lett.} {\bf 106} (2011)  161303}, \href{http://arxiv.org/abs/1011.6442}{{\tt
  arXiv:1011.6442 [gr-qc]}}.

\bibitem{Rama2012}
S.~K. Rama, ``{Remarks on Black Hole Evolution a la Firewalls and Fuzzballs},''
  \href{http://arxiv.org/abs/1211.5645}{{\tt arXiv:1211.5645}}.

\bibitem{Almheiri:2013wka}
A.~Almheiri and J.~Sully, ``{An Uneventful Horizon in Two Dimensions},''
  \href{http://dx.doi.org/10.1007/JHEP02(2014)108}{{\em JHEP} {\bf 02} (2014)
  108}, \href{http://arxiv.org/abs/1307.8149}{{\tt arXiv:1307.8149 [hep-th]}}.
  
\bibitem{Chen:2015}
P.~Chen, Y.~C.~Ong, and D.~Yeom, 
``Black Hole Remnants and the Information Loss Paradox,"
{\em Phys. Rept.} {\bf 603} (2015) 1, \href{http://arxiv.org/abs/1412.8366}{{\tt arXiv:1412.8366 [gr-qc]}}
  
\bibitem{Malafarina:2017}
D.~Malafarina,
``Classical collapse to black holes and quantum bounces: A review,"
{\em Universe} {\bf 3} (2017) 48, \href{http://arxiv.org/abs/1703.04138}{{\tt arXiv:1703.04138 [gr-qc]}}

  
\bibitem{Banks1992}
T.~Banks, A.~Dabholkar, M.~R. Douglas, and M.~O'Loughlin, ``{Are horned
  particles the end point of Hawking evaporation?},'' {\em Physical Review D}
  {\bf 45} (1992) no.~10, 3607--3616, \href{http://arxiv.org/abs/9201061}{{\tt
  arXiv:9201061 [hep-th]}}.

\bibitem{Giddings1992c}
S.~B. Giddings and A.~Strominger, ``{Dynamics of extremal black holes},''
  \href{http://dx.doi.org/10.1103/PhysRevD.46.627}{{\em Physical Review D} {\bf
  46} (1992) no.~2, 627--637}, \href{http://arxiv.org/abs/9202004}{{\tt
  arXiv:9202004 [hep-th]}}. \url{http://arxiv.org/abs/hep-th/9202004}.

\bibitem{Banks1993b}
T.~Banks, M.~O'Loughlin, and A.~Strominger, ``{Black hole remnants and the
  information puzzle},'' \href{http://dx.doi.org/10.1103/PhysRevD.47.4476}{{\em
  Physical Review D} {\bf 47} (1993) no.~10, 4476--4482},
  \href{http://arxiv.org/abs/9211030}{{\tt arXiv:9211030 [hep-th]}}.
  \url{http://arxiv.org/abs/hep-th/9211030}.

\bibitem{Giddings1994}
S.~B. Giddings, ``{Constraints on black hole remnants},'' {\em Physical Review
  D} {\bf 49} (1994) no.~2, 947--957, \href{http://arxiv.org/abs/9304027}{{\tt
  arXiv:9304027 [hep-th]}}.

\bibitem{Marolf2017}
D.~Marolf, ``{The Black Hole information problem: past, present, and future},''
  {\em Reports on Progress in Physics} {\bf 80} (2017)  092001,
  \href{http://arxiv.org/abs/1703.02143}{{\tt arXiv:1703.02143}}.

\bibitem{Bianchi:2014bma}
E.~Bianchi, T.~{De Lorenzo}, and M.~Smerlak, ``{Entanglement entropy production
  in gravitational collapse: covariant regularization and solvable models},''
  \href{http://dx.doi.org/10.1007/JHEP06(2015)180}{{\em Journal of High Energy
  Physics} {\bf 2015} (2015) no.~6, },
  \href{http://arxiv.org/abs/1409.0144}{{\tt arXiv:1409.0144}}.

\bibitem{Frolov2012}
V.~Frolov and I.~Novikov, {\em {Black Hole Physics: Basic Concepts and New
  Developments}}.
\newblock Springer, 2012.

\bibitem{Barrabes:1993}
C.~Barrab$\backslash$`es, P.~R. Brady, and E.~Poisson, ``{Death of white
  holes},'' {\em Phys. Rev. D} {\bf 47} (1993) no.~6, 2383----2387.

\bibitem{Poisson:1994}
A.~Ori and E.~Poisson, ``{Death of cosmological white holes},'' {\em Phys. Rev.
  D} {\bf 50} (1994) no.~10, 6150----6157.

\bibitem{Engelhardt2016}
N.~Engelhardt and G.~T. Horowitz, ``{Holographic consequences of a no
  transmission principle},'' {\em Physical Review D} {\bf 93} (2016)  026005,
  \href{http://arxiv.org/abs/1509.07509}{{\tt arXiv:1509.07509}}.

\bibitem{Fitzpatrick2016}
A.~L. Fitzpatrick, J.~Kaplan, D.~Li, and J.~Wang, ``{On information loss in
  AdS3/CFT2},'' {\em Journal of High Energy Physics} {\bf 2016} (2016)  ,
  \href{http://arxiv.org/abs/1603.08925}{{\tt arXiv:1603.08925}}.

\bibitem{Christodoulou2015}
M.~Christodoulou and C.~Rovelli, ``{How big is a black hole?},'' {\em Physical
  Review D} {\bf 91} (2015)  064046, \href{http://arxiv.org/abs/1411.2854}{{\tt
  arXiv:1411.2854}}.

\bibitem{Stanford2014}
D.~Stanford and L.~Susskind, ``{Complexity and shock wave geometries},'' {\em
  Physical Review D - Particles, Fields, Gravitation and Cosmology} {\bf 90}
  (2014) no.~12, , \href{http://arxiv.org/abs/1406.2678}{{\tt
  arXiv:1406.2678}}.

\bibitem{Perez2015}
A.~Perez, ``{No firewalls in quantum gravity: the role of discreteness of
  quantum geometry in resolving the information loss paradox},'' {\em Classical
  and Quantum Gravity} {\bf 32} (2015)  084001,
  \href{http://arxiv.org/abs/1410.7062}{{\tt arXiv:1410.7062}}.

\bibitem{Ori:2016}
A.~Ori, ``{Firewall or smooth horizon?},'' {\em General Relativity and
  Gravitation} {\bf 48} (2016) no.~1, 1--13,
  \href{http://arxiv.org/abs/arXiv:1208.6480v1}{{\tt arXiv:arXiv:1208.6480v1}}.

\bibitem{AshtekarILQGS:2015}
A.~Ashtekar, ``{The Issue of Information Loss: Current Status},'' in {\em
  International Loop Quantum Gravity Seminar, February 9th (2016)}.
\newblock 2016.

\bibitem{Susskind:2018fmx}
L. Susskind,
``Black Holes and Complexity Classes," \href{http://arxiv.org/abs/1802.02175}{{\tt arXiv:1802.02175}}.

\bibitem{Bengtsson2015}
I.~Bengtsson and E.~Jakobsson, ``{Black holes: Their large interiors},'' {\em
  Mod. Phys. Lett. A} {\bf 30} (2015)  1550103,
  \href{http://arxiv.org/abs/arXiv:1502.0190}{{\tt arXiv:arXiv:1502.0190}}.

\bibitem{Ong2015}
Y.~C. Ong, ``{Never Judge a Black Hole by Its Area},'' {\em Journal of
  Cosmology and Astroparticle Physics} {\bf 2015} (2015)  11,
  \href{http://arxiv.org/abs/1503.01092}{{\tt arXiv:1503.01092}}.

\bibitem{Wang2017}
S.-J. Wang, X.-X. Guo, and T.~Wang, ``{Maximal volume behind horizons without
  curvature singularity},'' \href{http://arxiv.org/abs/1702.05246}{{\tt
  arXiv:1702.05246}}.

\bibitem{Christodoulou2016a}
M.~Christodoulou and T.~{De Lorenzo}, ``{Volume inside old black holes},''
  \href{http://dx.doi.org/10.1103/PhysRevD.94.104002}{{\em Physical Review D}
  {\bf 94} (2016)  104002}, \href{http://arxiv.org/abs/1604.07222}{{\tt
  arXiv:1604.07222}}.
  
\bibitem{Ong:2015}
Y.~C.~Ong, 
 ``{The Persistence of the Large Volumes in Black Holes},"
{\em Gen. Rel. Grav.} {\bf 47} (2015) 88, \href{http://arxiv.org/abs/1503.08245}{{\tt arXiv:1503.08245 [gr-qc]}}

\bibitem{Rovelli2013d}
C.~Rovelli and F.~Vidotto, ``{Evidence for Maximal Acceleration and Singularity
  Resolution in Covariant Loop Quantum Gravity},'' {\em Physical Review
  Letters} {\bf 111} (2013) no.~9, 091303,
  \href{http://arxiv.org/abs/1307.3228}{{\tt arXiv:1307.3228}}.

\bibitem{Yonika:2017qgo}
A.~Yonika, G.~Khanna, and P.~Singh, ``{Von-Neumann Stability and Singularity
  Resolution in Loop Quantized Schwarzschild Black Hole},'' {\em Class. Quant.
  Grav.} {\bf 35} (2018)  045007, \href{http://arxiv.org/abs/1709.06331}{{\tt
  arXiv:1709.06331}}.

\bibitem{DAmbrosio}
F.~D'Ambrosio and C.~Rovelli, ``{How Information Crosses Schwarzschild's Central Singularity},''   \href{http://arxiv.org/abs/1803.05015}{{\tt arXiv:1803.05015 [gr-qc]}}.

\bibitem{Synge1950}
Synge, ``{The Gravitational Field of a Particle},'' {\em Proc Irish Acad} {\bf
  A53} (1950)  83.

\bibitem{Peeters:1994jz}
K.~Peeters, C.~Schweigert, and J.~W. van Holten, ``{Extended geometry of black
  holes},'' \href{http://dx.doi.org/10.1088/0264-9381/12/1/015}{{\em Class.
  Quant. Grav.} {\bf 12} (1995)  173--180},
  \href{http://arxiv.org/abs/gr-qc/9407006}{{\tt arXiv:gr-qc/9407006 [gr-qc]}}.

\bibitem{Koslowski:2016hds}
T.~A. Koslowski, F.~Mercati, and D.~Sloan, ``{Through the Big Bang},''
  \href{http://arxiv.org/abs/1607.02460}{{\tt arXiv:1607.02460 [gr-qc]}}.

\bibitem{Haggard2016}
H.~M. Haggard and C.~Rovelli, ``{Quantum Gravity Effects around Sagittarius
  A*},'' {\em International Journal of Modern Physics D} {\bf 25} (2016)  1--5,
  \href{http://arxiv.org/abs/1607.00364}{{\tt arXiv:1607.00364}}.

\bibitem{Barcelo2014b}
C.~Barcel{\'{o}}, R.~Carballo-Rubio, L.~J. Garay, and G.~Jannes, ``{The
  lifetime problem of evaporating black holes: mutiny or resignation},''
  \href{http://arxiv.org/abs/1409.1501}{{\tt arXiv:1409.1501}}.

\bibitem{Barcelo2016}
C.~Barcel{\'{o}}, R.~Carballo-Rubio, and L.~J. Garay, ``{Black holes turn white
  fast, otherwise stay black: no half measures},'' {\em Journal of High Energy
  Physics} {\bf 2016} (2016)  1--21,
  \href{http://arxiv.org/abs/1511.00633}{{\tt arXiv:1511.00633}}.

\bibitem{Barrau2014c}
A.~Barrau and C.~Rovelli, ``{Planck star phenomenology},'' {\em Physics Letters
  B} {\bf 739} (2014)  405--409, \href{http://arxiv.org/abs/1404.5821}{{\tt
  arXiv:1404.5821}}.

\bibitem{Barrau2014b}
A.~Barrau, C.~Rovelli, and F.~Vidotto, ``{Fast radio bursts and white hole
  signals},'' {\em Physical Review D} {\bf 90} (2014)  127503,
  \href{http://arxiv.org/abs/1409.4031}{{\tt arXiv:1409.4031}}.

\bibitem{Barrau2015}
A.~Barrau, B.~Bolliet, F.~Vidotto, and C.~Weimer, ``{Phenomenology of bouncing
  black holes in quantum gravity: a closer look},''
  \href{http://arxiv.org/abs/1507.05424}{{\tt arXiv:1507.05424}}.

\bibitem{Barrau2016}
A.~Barrau, B.~Bolliet, M.~Schutten, and F.~Vidotto, ``{Bouncing black holes in
  quantum gravity and the Fermi gamma-ray excess},''
  \href{http://arxiv.org/abs/1606.08031}{{\tt arXiv:1606.08031}}.

\bibitem{Rovelli2017}
C.~Rovelli, ``{Planck stars as observational probes of quantum gravity},'' {\em
  Nature Astronomy} {\bf 1} (2017) no.~3, 0065.
  \url{http://www.nature.com/articles/s41550-017-0065}.

\bibitem{Retter2012}
A.~Retter and S.~Heller, ``{The revival of white holes as Small Bangs},'' {\em
  New Astronomy} {\bf 17} (2012)  73--75,
  \href{http://arxiv.org/abs/1105.2776}{{\tt arXiv:1105.2776}}.

\bibitem{Rovelli2017a}
C.~Rovelli, ``{Black holes have more states than those giving the
  Bekenstein-Hawking entropy: a simple argument},''
  \href{http://arxiv.org/abs/1710.00218}{{\tt arXiv:1710.00218}}.

\bibitem{Doeleman2009}
S.~S. Doeleman, E.~Agol, D.~Backer, F.~Baganoff, G.~C. Bower, A.~E. Broderick,
  A.~C. Fabian, V.~L. Fish, C.~Gammie, P.~Ho, M.~Honma, T.~Krichbaum, A.~Loeb,
  D.~Marrone, M.~J. Reid, A.~E.~E. Rogers, I.~Shapiro, P.~Strittmatter,
  R.~P.~J. Tilanus, J.~Weintroub, A.~Whitney, M.~Wright, and L.~Ziurys,
  ``{Imaging an Event Horizon: submm-VLBI of a Super Massive Black Hole},''
  {\em Astro2010: The Astronomy and Astrophysics Decadal Survey} {\bf 2010}
  (2009)  68, \href{http://arxiv.org/abs/0906.3899}{{\tt arXiv:0906.3899}}.

\bibitem{Page1993a}
D.~N. Page, ``{Information in black hole radiation},'' {\em Physical Review
  Letters} {\bf 71} (1993)  3743--3746,
  \href{http://arxiv.org/abs/0000135489}{{\tt arXiv:0000135489}}.

\bibitem{Almheiri2013}
A.~Almheiri, D.~Marolf, J.~Polchinski, and J.~Sully, ``{Black holes:
  Complementarity or firewalls?},'' {\em Journal of High Energy Physics} {\bf
  2013} (2013)  1--19, \href{http://arxiv.org/abs/1207.3123}{{\tt
  arXiv:1207.3123}}.

\bibitem{Giddings2013}
S.~B. Giddings, ``{Statistical physics of black holes as quantum-mechanical
  systems},'' {\em Physical Review D - Particles, Fields, Gravitation and
  Cosmology} {\bf 88} (2013) no.~10, ,
  \href{http://arxiv.org/abs/1308.3488}{{\tt arXiv:1308.3488}}.

\bibitem{Unruh2017}
W.~G. Unruh and R.~M. Wald, ``{Information Loss},'' {\em Reports on Progress in
  Physics} {\bf 80} (2017)  092002, \href{http://arxiv.org/abs/1703.02140}{{\tt
  arXiv:1703.02140}}.


\end{thebibliography}\endgroup

\end{document}